\documentclass[aps,prc,superscriptaddress,showpacs,amssymb,amsmath,amsfonts,twocolumn,floatfix]{revtex4}
\usepackage{graphicx}
\usepackage{amsmath}
\usepackage{amssymb}

\begin{document}

\include{defs}

\title{Partial-Wave Analysis of Nucleon-Nucleon Elastic Scattering Data}

\author{Ron L.\ Workman}
\affiliation{Institute for Nuclear Studies, Department of Physics,
        The George Washington University, Washington, D.C. 20052}
\author{William J.\ Briscoe}
\affiliation{Institute for Nuclear Studies, Department of Physics,
        The George Washington University, Washington, D.C. 20052}
\author{Igor I.\ Strakovsky}
\affiliation{Institute for Nuclear Studies, Department of Physics, 
	The George Washington University, Washington, D.C. 20052}

\date{\today}

%%%%%%%%%%%%%%%%%%%%%%%%%%%%%%%%%%%%%%%%%%%%%%%%%%%%%5
\begin{abstract}
Energy-dependent and single-energy fits to the existing nucleon-nucleon
database have been updated to incorporate recent measurements. The fits 
cover a region from threshold to 3~GeV, in the laboratory kinetic energy, 
for proton-proton scattering, with an upper limit of 1.3~GeV for 
neutron-proton scattering. Experiments carried out at the COSY-WASA and
COSY-ANKE facilities have had a significant impact on the partial-wave
solutions. Results are discussed in terms of both partial-wave and
direct reconstruction amplitudes. 
\end{abstract}


\maketitle

%%%%%%%%%%%%%%%%%%%%%%%%%%%%%%%%%%%%%%%%%%%%%%%%%%%%%5
\section{Introduction}
\label{sec:intro} 

The nucleon-nucleon interaction is fundamental to the understanding
of nuclear physics as a whole. The elastic scattering of protons and 
neutrons, at low to medium energies, can be analyzed in terms of a
partial-wave expansion, yielding amplitudes that are valuable for
comparison to models~\cite{models} and also as input to more complex 
processes, such as proton-deuteron scattering~\cite{pd_scatt} and 
deuteron electro-disintegration~\cite{van_orden}.
Fit results have also been used in Monte Carlo simulations and 
incorporated into geometry and tracking of the Geant4 simulation~\cite{MC} 
software, a basic tool for low, medium, and high energy groups. 

In nucleon-nucleon elastic scattering, as the initial and final state 
particles all have spin 1/2, many independent observables 
are available~\cite{formal} 
and have been measured over decades of systematic studies at ANL,  
COSY, IUCF, LAMPF, PSI, SATURN, TRIUMF, TUNL and other facilities. This 
has allowed a direct scattering amplitude 
reconstruction (DAR)~\cite{Bystricky98, Ball98} at 
a number of energies and angles. We compare our results with these 
reconstructions below. 

Our last published analysis~\cite{SP07} (SP07) of the nucleon-nucleon 
database was completed in 2007 and largely motivated by data provided
by the EDDA Collaboration~\cite{EDDA}, with some data also coming from 
the SATURNE~II and PNPI groups. The most recent measurements have been 
fewer in number but greater in effect. This is true for both $pp$ and
$np$ scattering.

Recent COSY-WASA measurements~\cite{WASA,dibaryon_GW} of the $np$ 
scattering observable, $A_y$, have shown a sharp energy dependence, over 
a narrow energy range, when combined with the trend displayed by existing 
lower-energy data~\cite{old_Ay}. This behavior was not predicted by the 
SP07 analysis, nor was it seen in previous fits~\cite{SM97,SP00,SM94,FA91}. 
In a revised fit~\cite{WASA}, this narrow structure was reproduced with 
the generation of a pole, at $[(2380\pm 10) - i(40\pm 5)]$~MeV in the 
coupled $^3D_3$--$^3G_3$ partial waves, closely related to the resonance 
mass and width ($M\approx 2380$~MeV , $\Gamma \approx 70$~MeV), deduced 
from WASA analyses of related two-pion production measurements~\cite{2pi}. 

We subsequently compared the predictions of fits to $np$ scattering data, 
with and without a pole, to search for other potentially sensitive 
observables~\cite{dibaryon_GW}. Experimental tests are limited by the
existing $np$ database, which supports a partial wave analysis (PWA)
only up to a laboratory kinetic energy of 1.3~GeV. No single measurement 
is capable of extending the range of PWAs. A programmatic 
study would be required and this is unlikely to occur in the near future. 

For $pp$ elastic scattering, measurements of near-forward differential 
cross sections~\cite{mc16} and the polarization quantity~\cite{ba14}, $A_y$, 
from the COSY-ANKE Collaboration, were poorly predicted by the SP07
fit. The simple inclusion of these data in the full database did not 
generally result in a significant improvement. Below we discuss the 
fitting strategy and its connection to recent (SP07)~\cite{SP07} and 
older (SM97)~\cite{SM97} PWA results.
  
%%%%%%%%%%%%%%%%%%%%%%%%%%%%%%%%%%%%%%%%%%%%%%%%
\section{Fits to Data}
\label{sec:FITS}

The SP07 analysis is accessible via the Scattering 
Analysis Interactive Database (SAID) website~\cite{SAID}. The fit form is 
based on the product of exchange and production S-matrices. The exchange 
piece is parametrized via a K-matrix and contains a one-pion exchange term 
plus a sum over expansion bases containing a left-hand cut. The production 
piece is parametrized in terms of a Chew-Mandelstam K-matrix and allows 
for the production of an inelastic channel. For the isovector waves, the 
inelasticity is assumed to be dominated by the $N \Delta$ channel. For 
isoscalar waves, there is an effective second channel. The formalism for 
both spin-coupled and uncoupled waves is explained in detail in 
Refs.~\cite{SM97,formalism}.

The $\chi^2$ fit to data is carried out, using the form
%%%%%%%%%%%%%%%%%%%%%%%%%%%%%%%%%%%%%%%%%%%%%%%%%%%%%5
\begin{equation}
	\chi^2 = \sum_i \left( {{N \Theta_i - \Theta_i^{\rm exp} }\over 
	{\epsilon_i}} \right)^2 + \left( {{N-1}\over{\epsilon_N}} \right)^2 ,
	\label{chi0}
\end{equation}
%%%%%%%%%%%%%%%%%%%%%%%%%%%%%%%%%%%%%%%%%%%%%%%%%%%%%5
where $\Theta_i^{\rm exp}$ is an experimental point in an angular 
distribution, with associated statistical error $\epsilon_i$, and $\Theta_i$ 
is the fit value. Here the overall systematic error, $\epsilon_N$, is used 
to weight an additional $\chi^2$ penalty term due to the renormalization of
angular distributions in the fit by the factor $N$.

The partial-wave solution (SP07) was determined through a fit to approximately
25 thousand $pp$ data (to 3~GeV) and 13 thousand $np$ data (to 1.3~GeV). 
Compared to the recent $np$~\cite{WASA} and $pp$~\cite{ba14} $A_y$ measurements, 
SP07 predicted a different forward-angle behavior, for $pp$ data, and a 
different shape at the energy corresponding to the proposed WASA dibaryon, for 
$np$ data. 
SP07 was able to reproduce the angular behavior of
the new and very precise forward $pp$ cross sections~\cite{mc16}, but 
the resulting renormalization factors ($N$), for several energies, were outside 
the range expected from the quoted overall systematic errors. 

Different strategies were used to accommodate the new $np$ and $pp$ scattering
data.  For the $pp$ $A_y$ observable, an improved fit was achieved by more 
heavily weighting the new data, by a factor of 4, in the fit. Conversely, for 
the cross sections, the second term in Eq.~\ref{chi0} was increased in weight 
until the fitted renormalization factors deviated from unity by an amount 
consistent with the quoted systematic errors. Weighting was also used to 
study the influence of the COSY-WASA $np$ $A_y$ data, as has been described 
in Refs.~\cite{WASA,dibaryon_GW}. 

\begin{figure}[th]
\centerline{
\includegraphics[height=0.3\textwidth, angle=90]{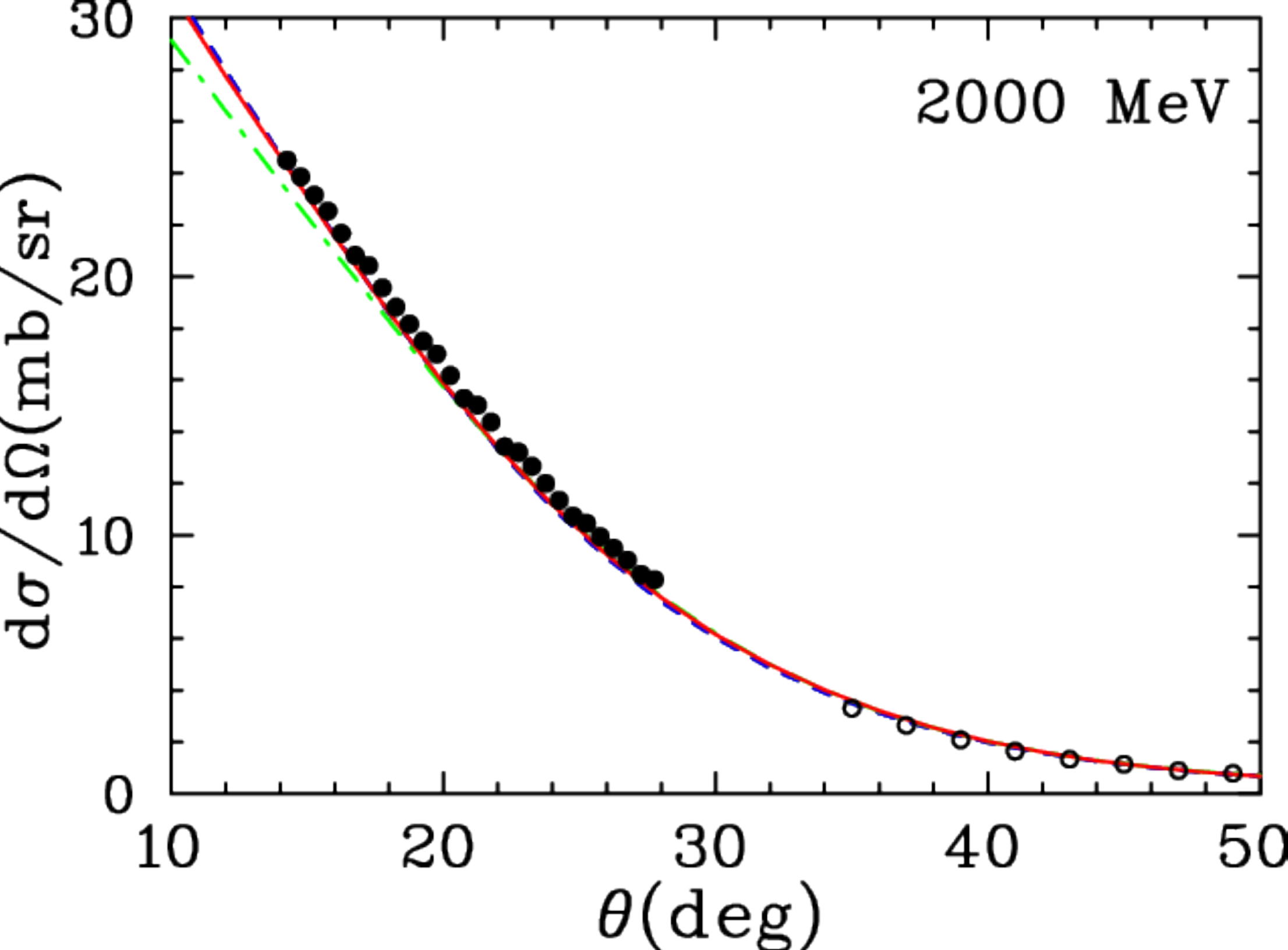}}
\centerline{
\includegraphics[height=0.3\textwidth, angle=90]{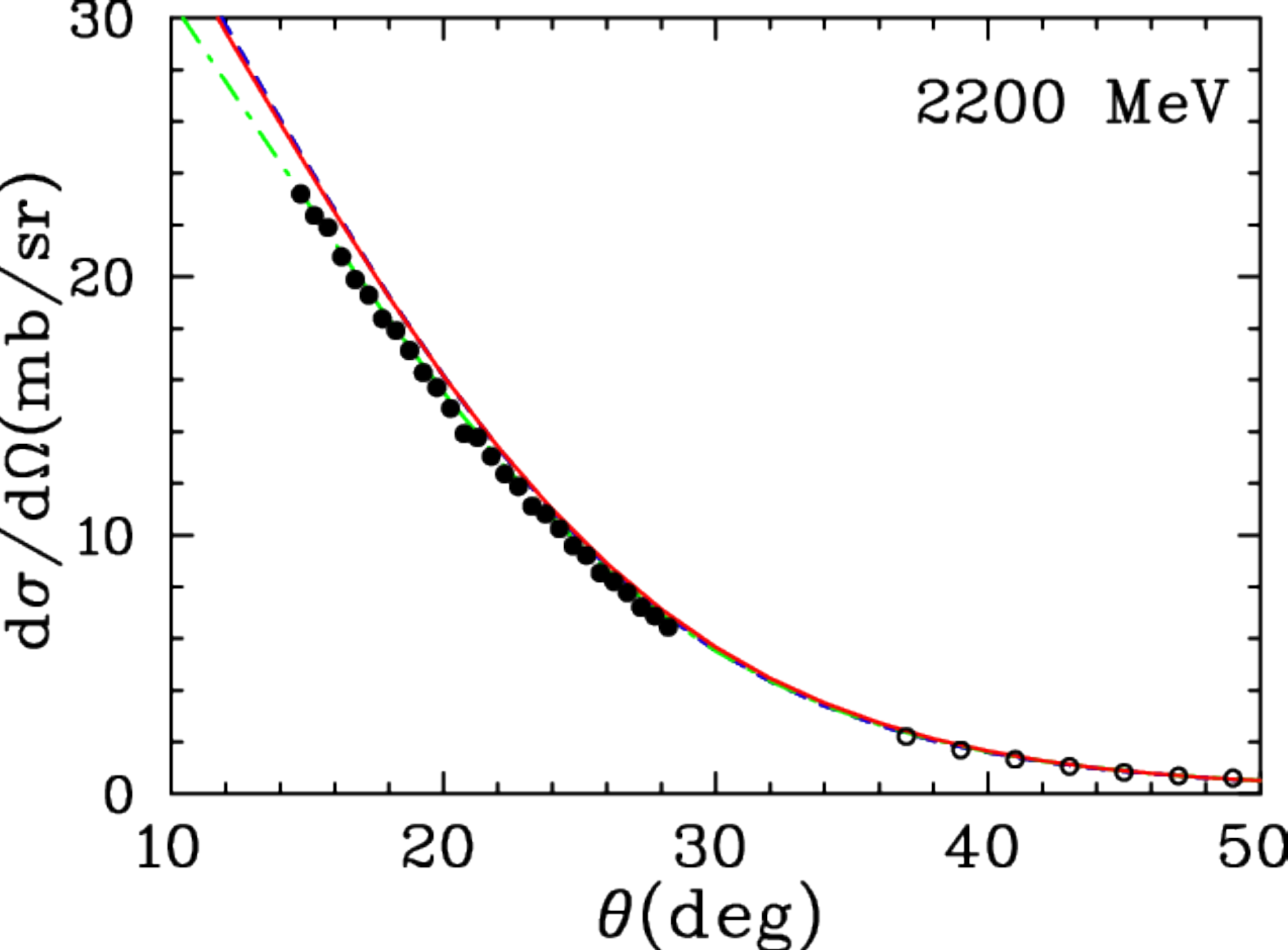}}
\centerline{
\includegraphics[height=0.3\textwidth, angle=90]{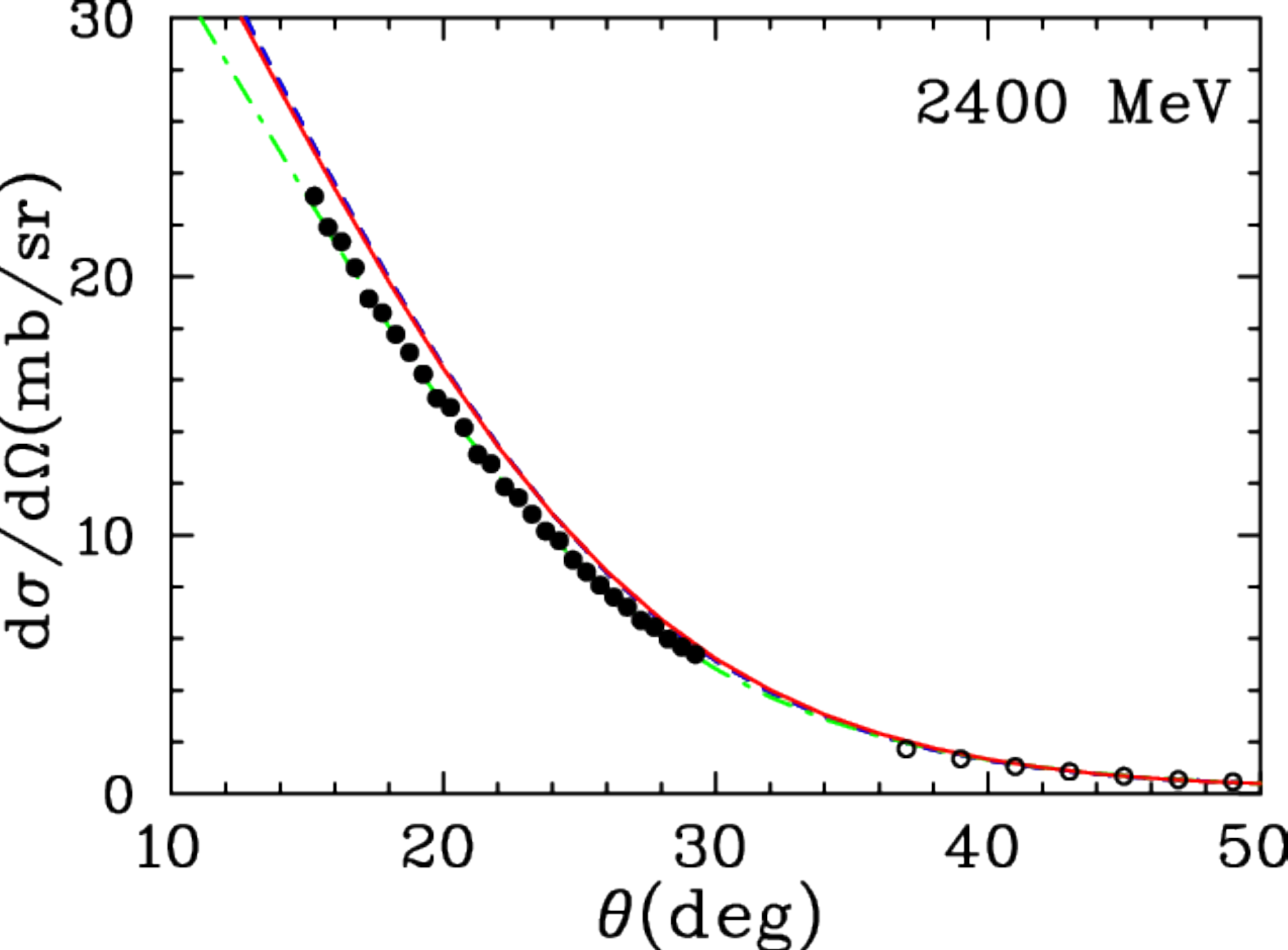}}
\centerline{
\includegraphics[height=0.3\textwidth, angle=90]{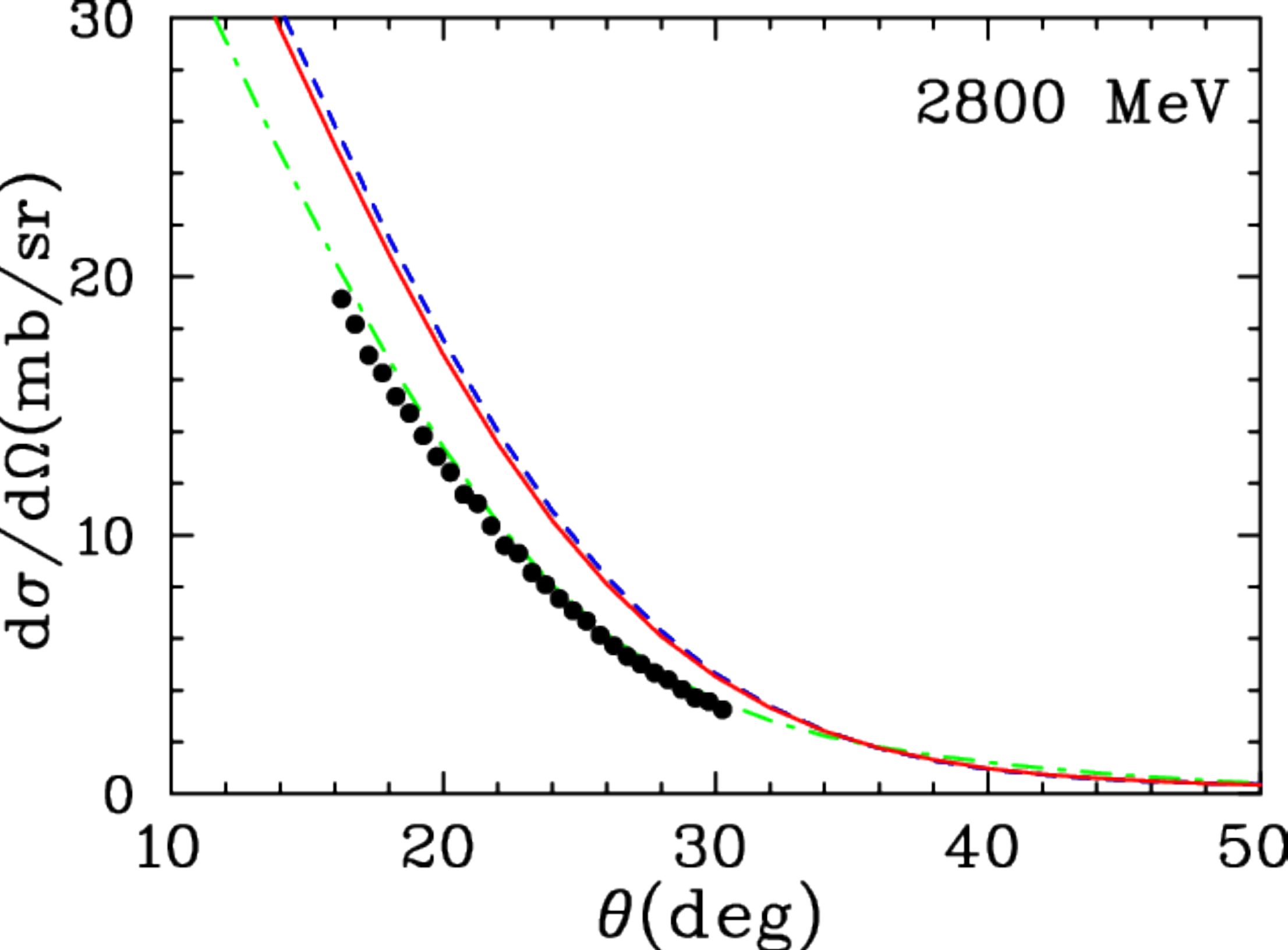}}
\vspace{3mm}
\caption{(Color online) Cross sections for $pp$ elastic scattering.
        Data from the SAID database~\protect\cite{SAID}. Recent
        COSY data plotted as solid black circles; older
        COSY data within $\pm 4$~MeV shown as open black 
	circles.  Red solid (green long dot-dashed) lines, 
	correspond to the recent SM16 (WF16) solution.  The 
	previous SP07~\protect\cite{SP07} solution is plotted with 
	blue dashed lines. \label{fig:g1}}
\end{figure}

In Fig.~\ref{fig:g1}, we plot the fits SP07, an unweighted fit (SM16), and 
a weighted fit (WF16) to the new forward cross section data~\cite{mc16}. 
The fit quality for larger-angle COSY measurements~\cite{al04}, which were 
included in the SP07 fit, is displayed as well. In the comparison, the 
fitted normalization factor has not been applied in order to show how 
large a factor is required in an unweighted fit, or the SP07 prediction, 
at the highest energies. 
%%%%%%%%%%%%%%%%%%%%%%%%%%%%%%%%%%%%%%%%%%%%%%%%%%%%
%%%%%%%%%%%%%%%%%%%%%%%%%%%%%%%%%%%%%%%%%%%%%%%%%%%%

\begin{figure}[th]
\centerline{
\includegraphics[height=0.3\textwidth, angle=90]{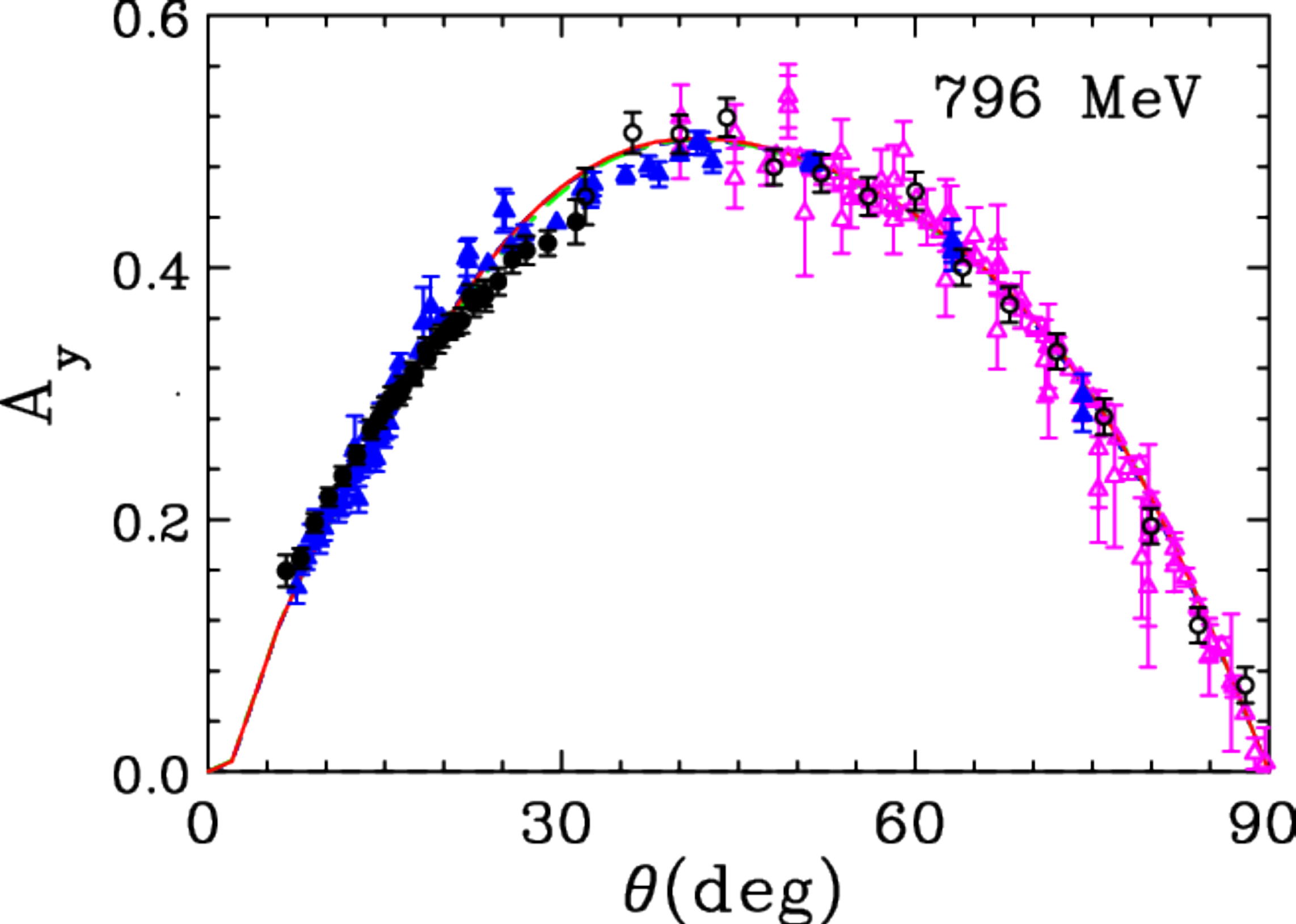}}
\centerline{
\includegraphics[height=0.3\textwidth, angle=90]{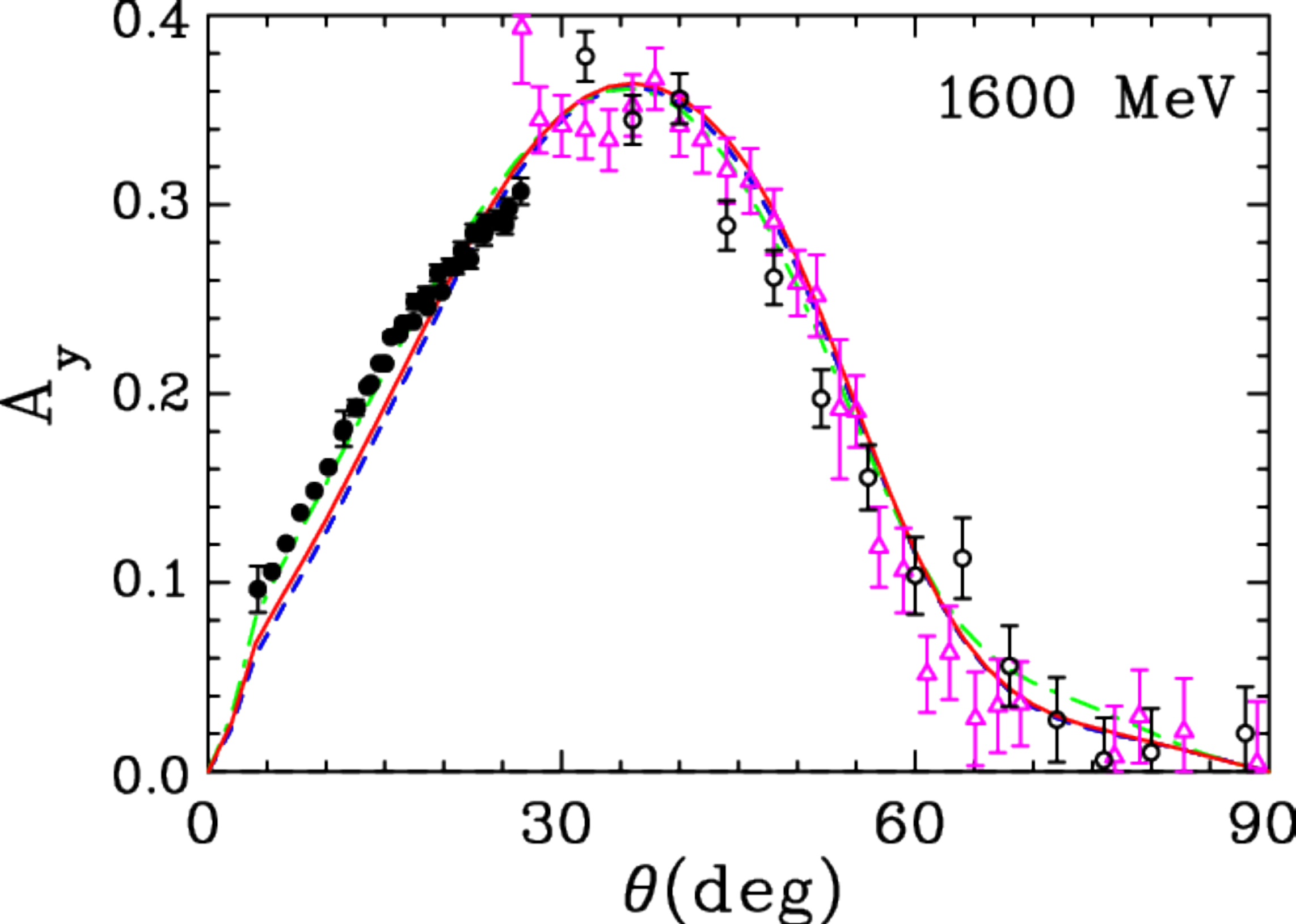}}
\centerline{
\includegraphics[height=0.3\textwidth, angle=90]{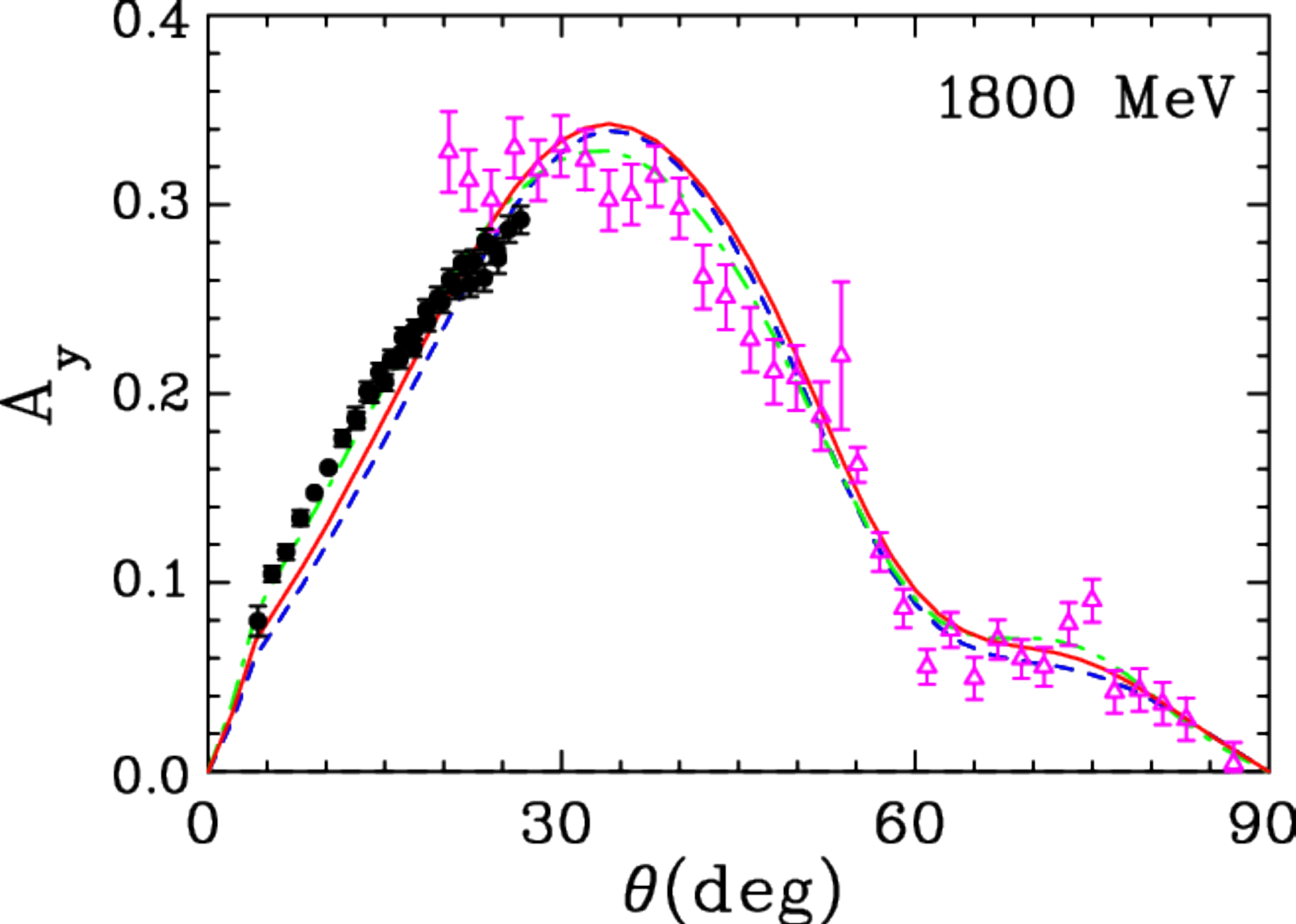}}
\centerline{
\includegraphics[height=0.3\textwidth, angle=90]{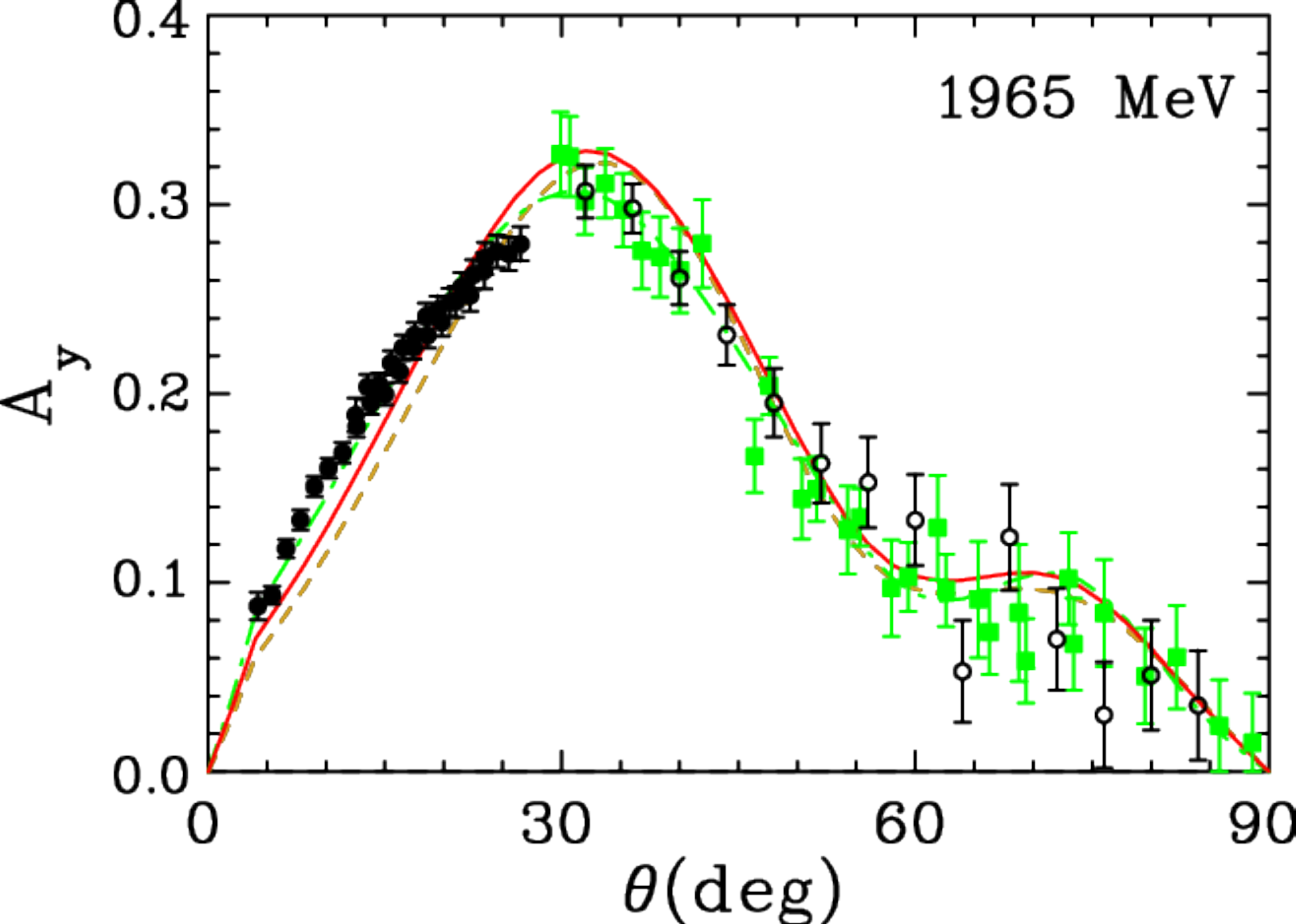}}
\vspace{3mm}
\caption{(Color online) Polarized observable $A_y$ for $pp$
        elastic scattering. Recent COSY data plotted as solid
        black circles. Data within $\pm 8$~MeV
        from the SAID database~\protect\cite{SAID}
        (COSY, SATURN, LAMPF, and ANL) shown as
        open black circles, open magenta triangles,
        solid blue triangles, and solid green
        squares.
        Notation for energy dependent fits as in
        Fig.~\protect\ref{fig:g1}. \label{fig:g2}}
\end{figure}

In Fig.~\ref{fig:g2}, the above three partial-wave solutions are 
compared to the new $pp$ $A_y$ data~\cite{ba14} and other data sets 
covering the full angular range. In contrast to Fig.~1, the data sets 
have all been modified according to the renormalization factors found 
in the weighted fit WF16. Here we see that the new $A_y$ measurements 
cover an angular range not previously measured, are much more precise 
than earlier measurements and, in some cases, differ from those earlier 
measurements at angles where there is overlap. 
%%%%%%%%%%%%%%%%%%%%%%%%%%%%%%%%%%%%%%%%%%%%%%%%%%%%
%%%%%%%%%%%%%%%%%%%%%%%%%%%%%%%%%%%%%%%%%%%%%%%%%%%%%%

\begin{figure}[th]
\centerline{
\includegraphics[height=0.3\textwidth, angle=90]{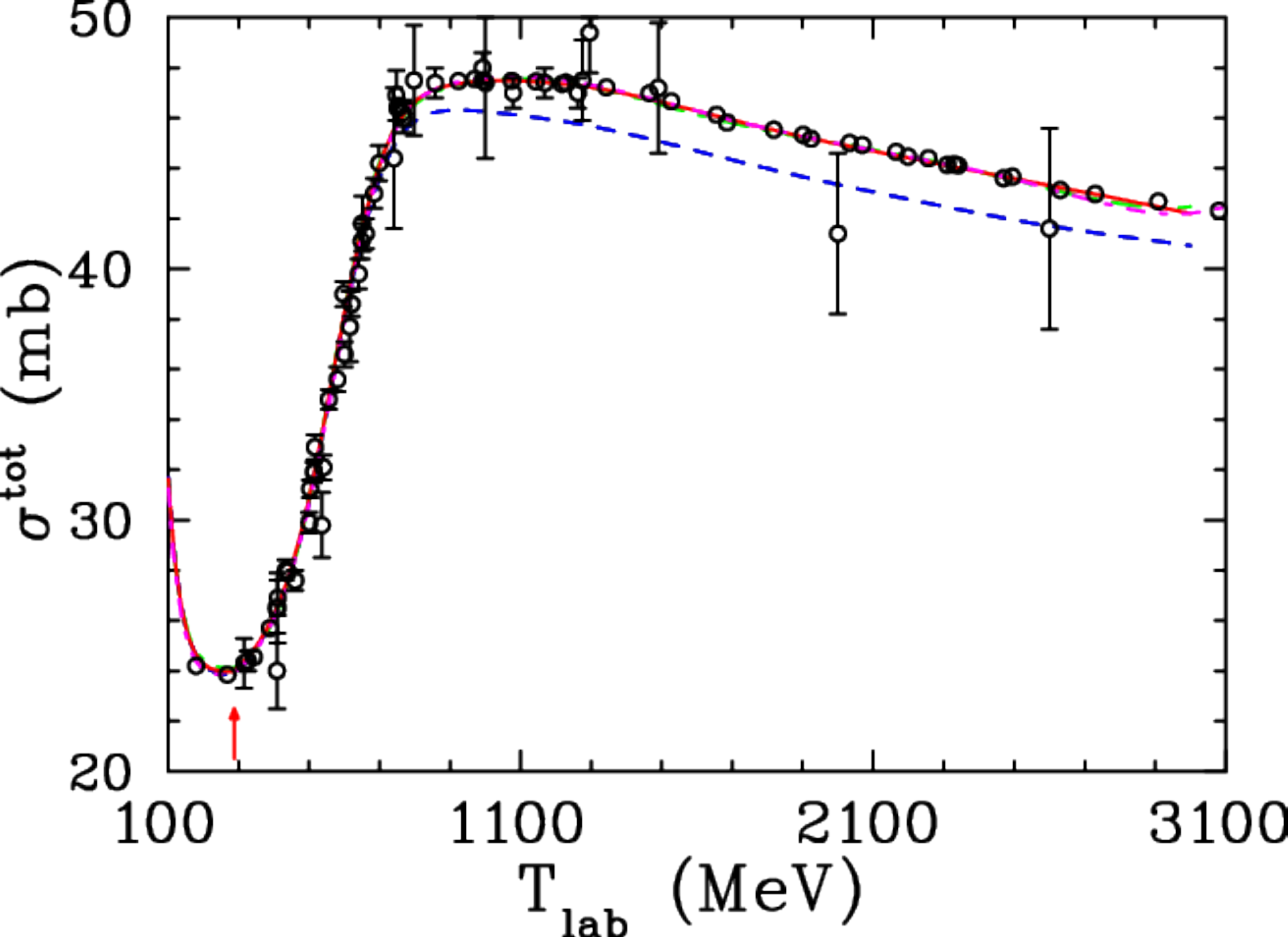}}
\centerline{
\includegraphics[height=0.3\textwidth, angle=90]{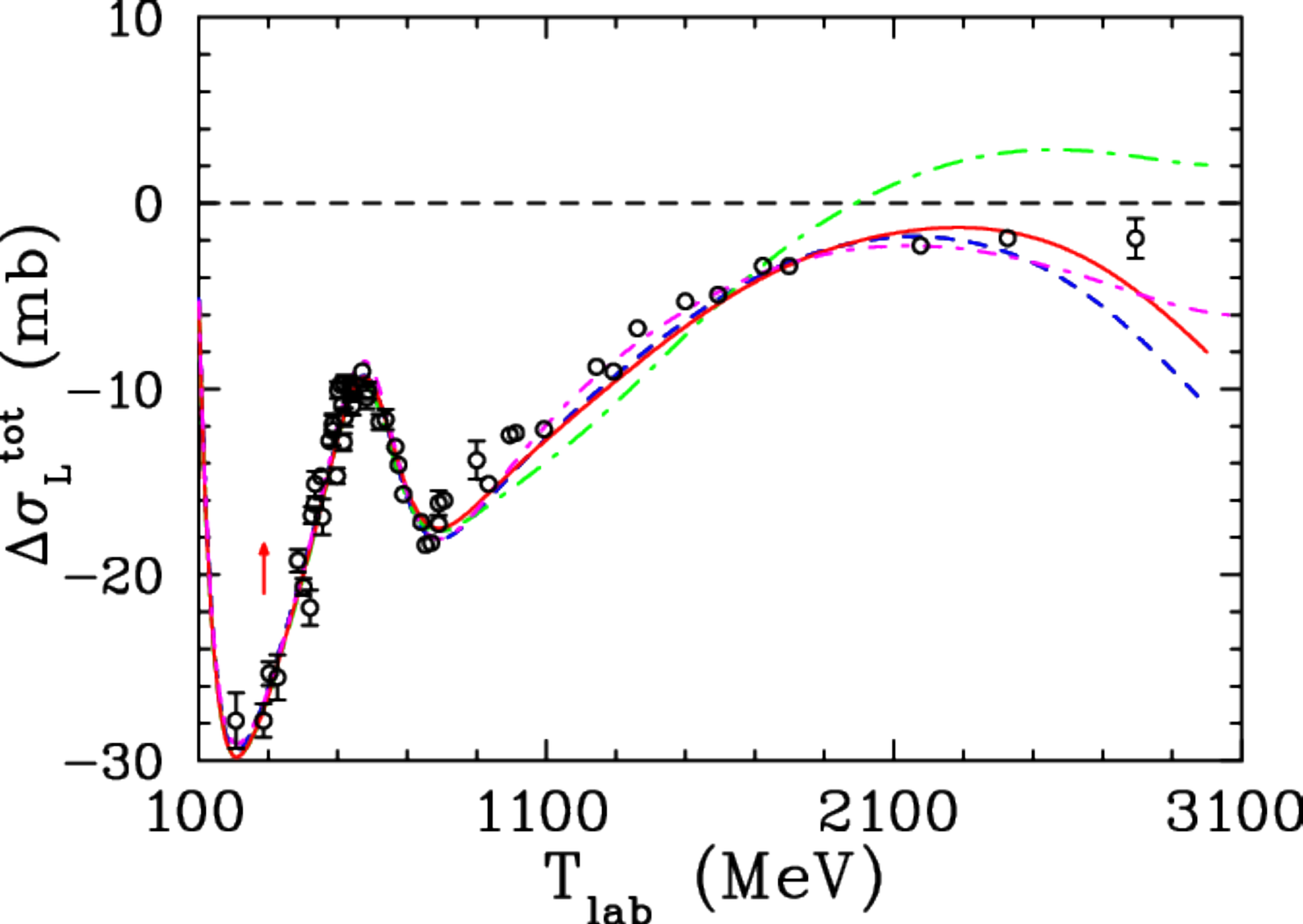}}
\centerline{
\includegraphics[height=0.3\textwidth, angle=90]{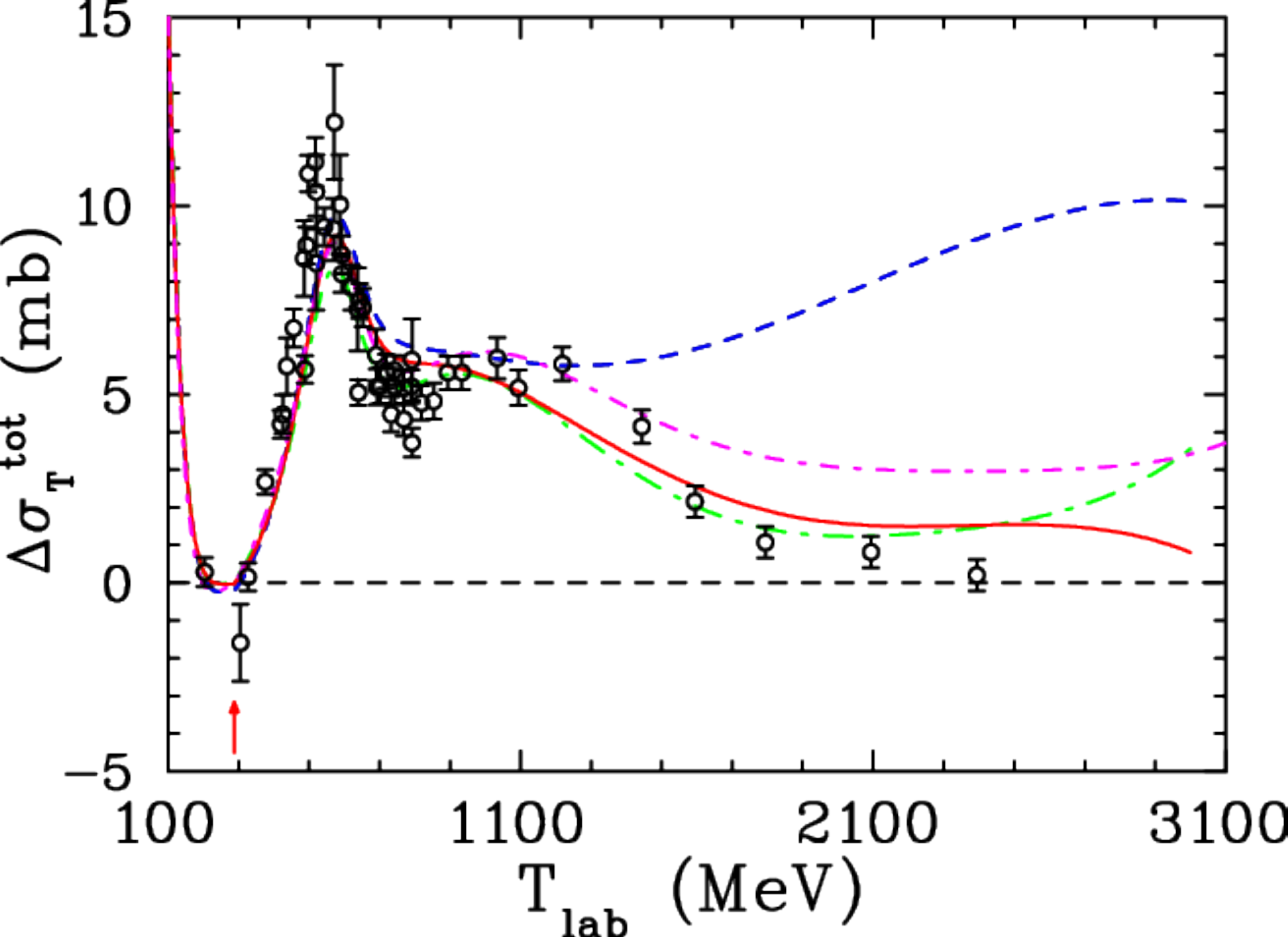}}
\centerline{
\includegraphics[height=0.3\textwidth, angle=90]{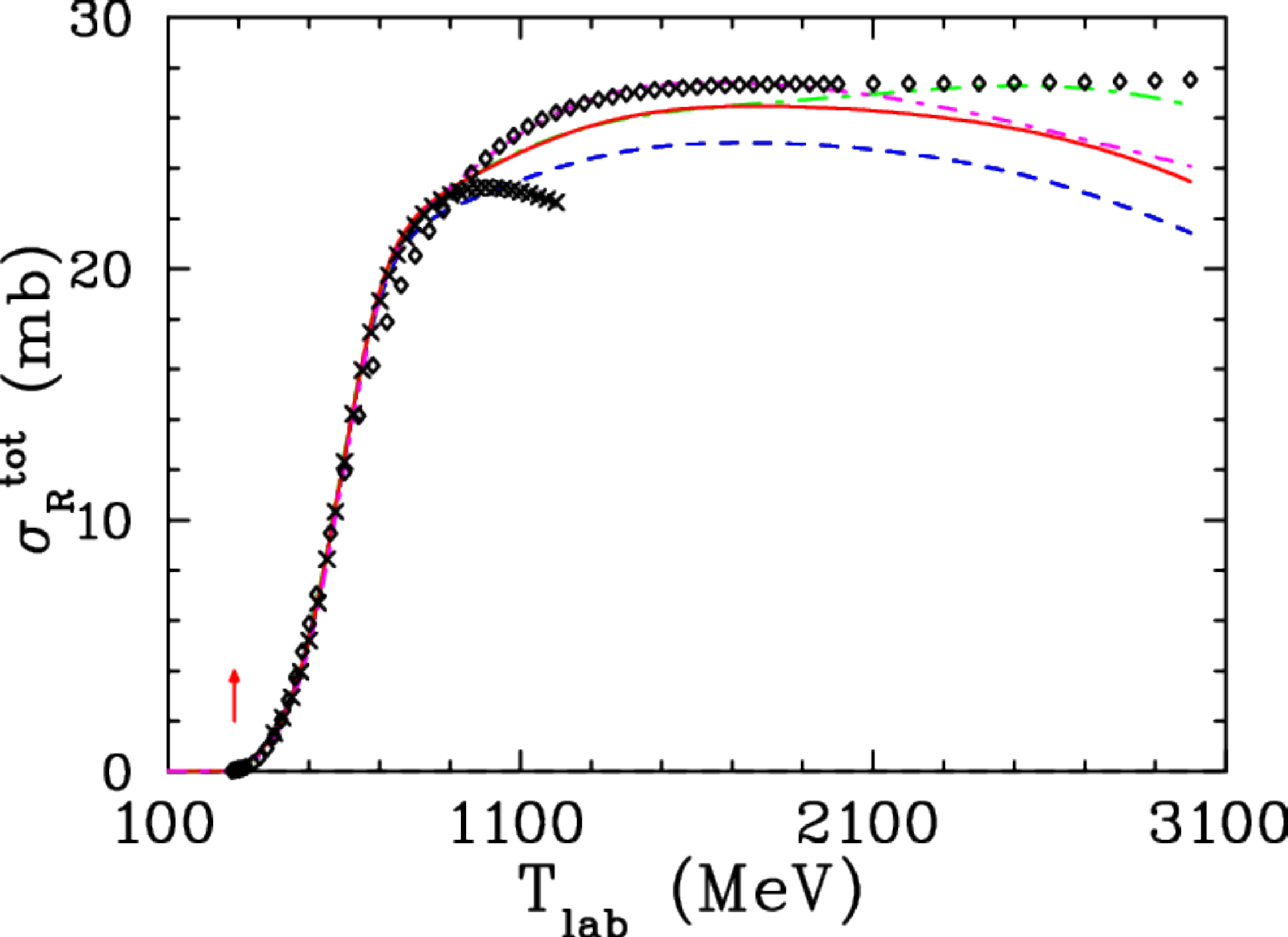}}
\vspace{3mm}
\caption{(Color online) Total unpolarized and polarized cross
        sections for $pp$ scattering. Quasi-data for total 
	reaction cross sections from Refs.~\cite{ve82} and \cite{by78}
	are given by black crosses
	and open diamonds, respectively. 
	Pion production threshold shown as a vertical red arrow.
        Notation for energy dependent fits as in
        Fig.~\protect\ref{fig:g1}. Additionally, the older SM97
        solution~\protect\cite{SM97} is plotted as magenta 
	short dot-dashed lines. \label{fig:g3}}
\end{figure}

In the revised solutions SM16 and WF16, the fit quality for total cross 
sections was also addressed. In Fig.~\ref{fig:g3}, we plot the total 
cross section ($\sigma^{\rm tot}$), the total reaction cross section 
($\sigma_{\rm R}^{\rm tot}$), the difference between total cross sections in 
pure transverse-spin states 
\begin{equation}
\Delta \sigma_T^{\rm tot} = \sigma^{tot}(\uparrow\downarrow) - 
\sigma^{tot}(\uparrow\uparrow) ,
\end{equation} 
and the difference between total cross sections in pure longitudinal-spin 
states 
\begin{equation}
\Delta \sigma_L^{\rm tot} = \sigma^{tot}(\leftarrow\rightarrow) - 
\sigma^{tot}(\rightarrow\rightarrow) .
\end{equation} 
%%%%%%%%%%%%%%%%%%%%%%%%%%%%%%%%%%%%%%%%%%%%%%%%%%%%%%
%%%%%%%%%%%%%%%%%%%%%%%%%%%%%%%%%%%%%%%%%%%%%%%%%%%%%%

Several comments are in order. The total reaction cross section 
~\cite{ve82,by78} is not
a pure measurement and has not been included in these fits. It is, 
however, qualitatively well predicted. The quantity $\Delta 
\sigma_T^{\rm tot}$ is not well reproduced by the SP07 solution and was 
given added weight in the SM16 and WF16 fits. In contrast, $\Delta 
\sigma_L^{\rm tot}$ is reasonably well represented by all the displayed 
fits. The total cross section $\sigma^{\rm tot}$ dataset includes some 
extremely precise measurements~\cite{bu66} (with statistical errors of 
the order of 0.1\%) which can dominate the $\chi^2$ in a fit. In the 
SP07 fit, subsets of energies in the measurement of Ref.~\cite{bu66} 
were allowed to have a common renormalization, resulting in a curve 
systematically below these data. In the present fit, this 
renormalization freedom was removed. As a result, the fits SM16 and 
WF16 are nearly identical for this quantity. For comparison purposes, 
the older fit SM97~\cite{SM97} has been plotted as well. Below we will 
see that the quality of fit to total cross section quantities is 
reflected in a variability for the $^1S_0$ partial wave amplitude.

\begin{figure}[th]
\centerline{
\includegraphics[height=0.3\textwidth, angle=90]{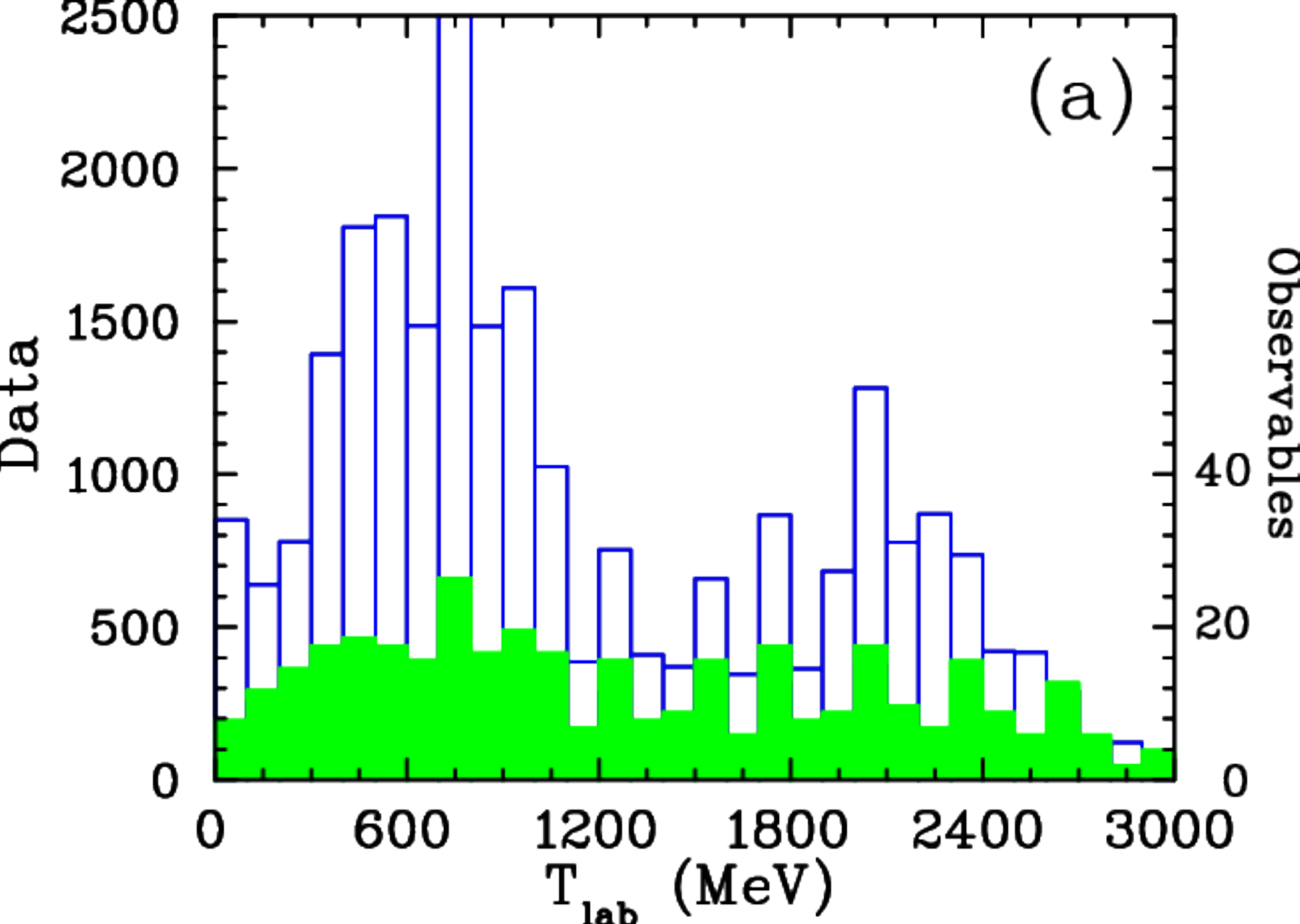}}
\centerline{
\includegraphics[height=0.3\textwidth, angle=90]{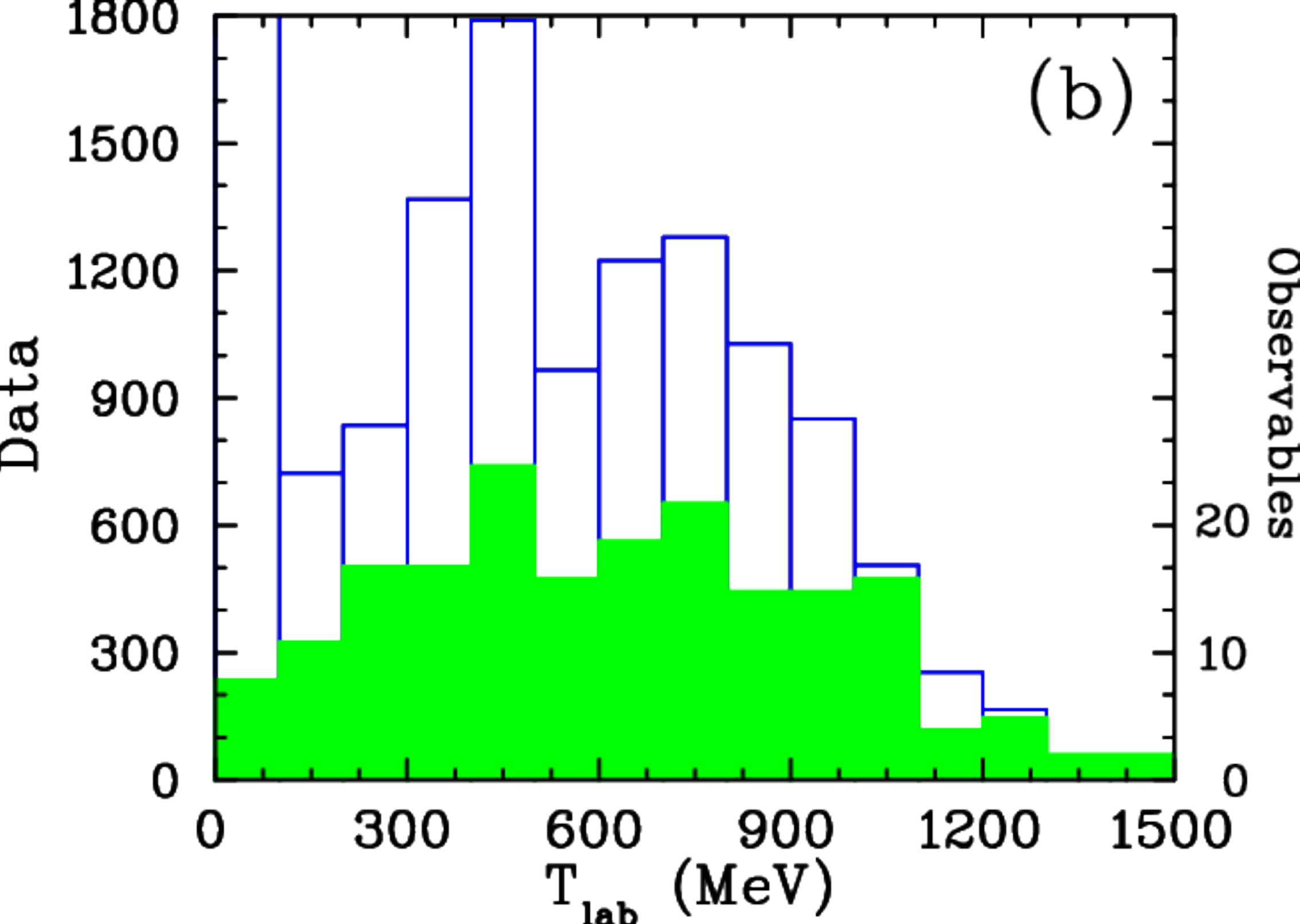}}
\vspace{3mm}
\caption{(Color online) Histograms (a) for $pp$ scattering data (open
        blue bars) and observable types (filled green bars) and (b) for
        $np$ data in the same notation.
	 \label{fig:g4}}
\end{figure}

\begin{figure}[th]
\centerline{
\includegraphics[height=0.3\textwidth, angle=90]{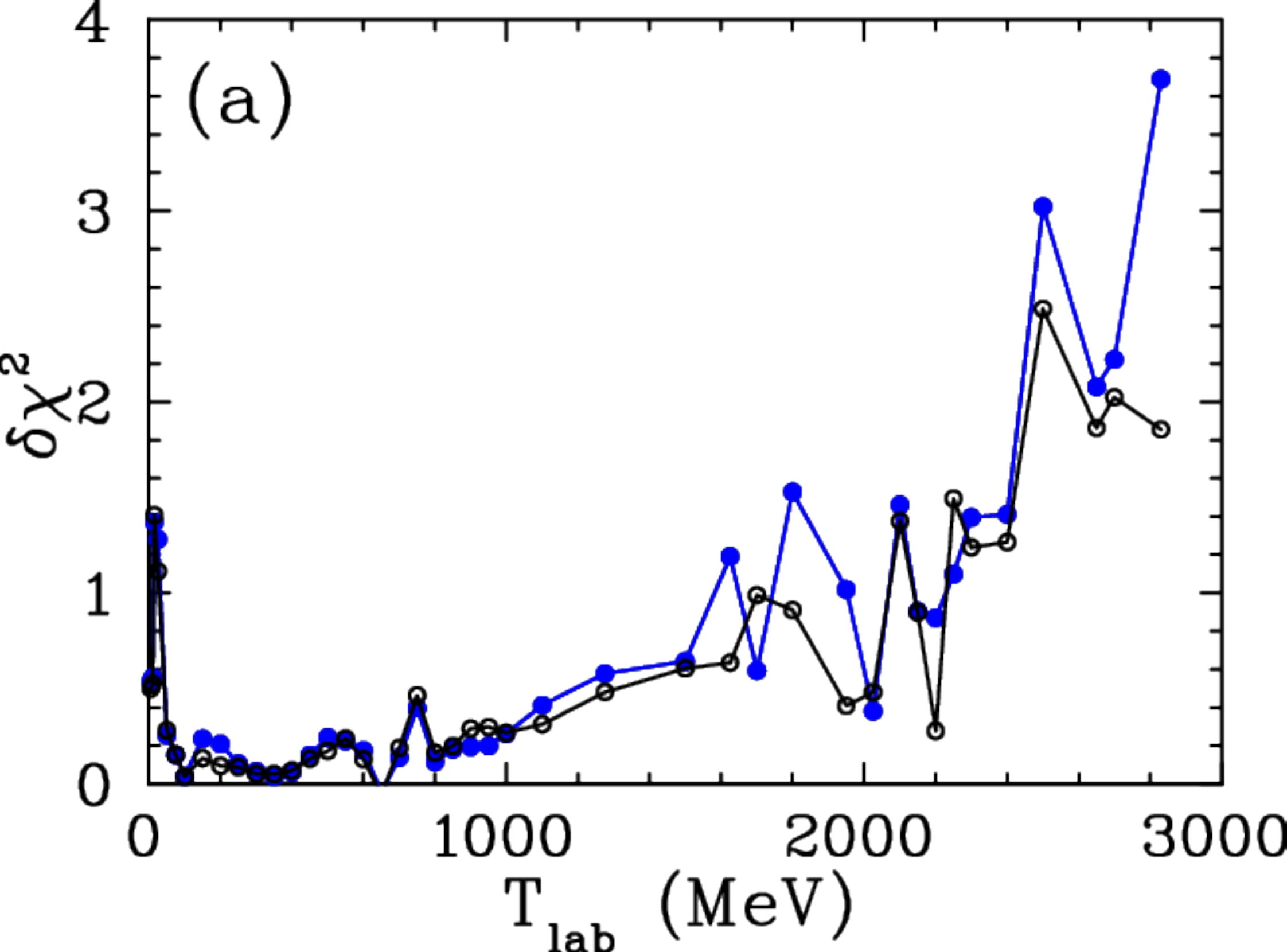}}
\centerline{
\includegraphics[height=0.3\textwidth, angle=90]{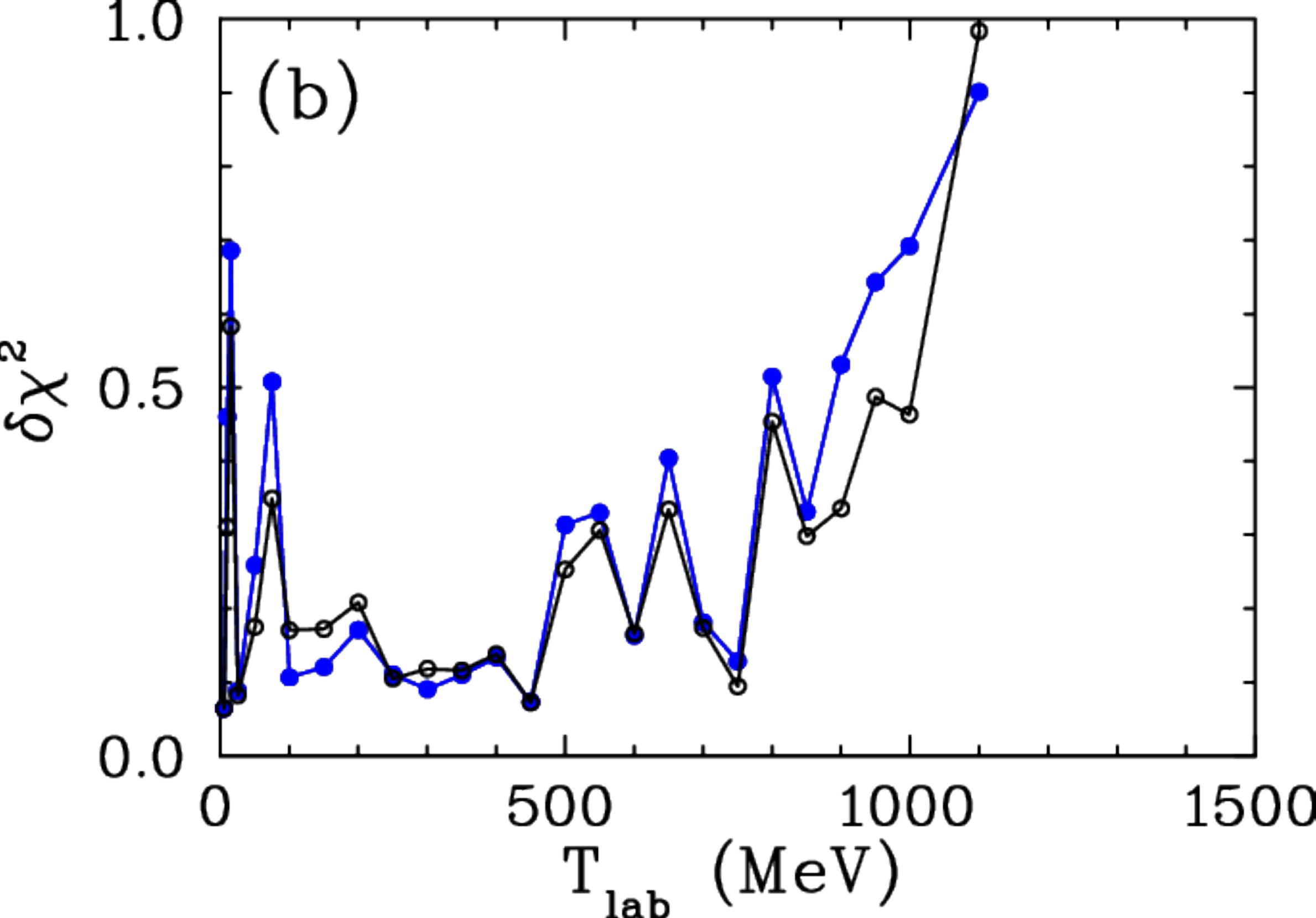}}
\vspace{3mm}
\caption{(Color online) Comparison of the SES and energy dependent (ED)
         fits in terms of $\delta\chi^2 = [\chi^2(ED) -
        \chi^2(SES)]/{\rm data}$  
	 for (a) $pp$ and (b) $np$ data. The SM16 and SP07 ED fits are
	displayed as blue solid and black open symbols, respectively. 
	\label{fig:g5}}
\end{figure}

In addition to global fits to all data from threshold to 3~GeV and 1.3~GeV,
in $T_{\rm lab}$, for $pp$ and $np$ scattering data, respectively, a set of 
single-energy solutions (SES) was also generated. Here, data within narrow 
energy bins are searched, starting from the global fit, without the 
constraint of energy-to-energy smoothness. This allows a search for 
systematic deviations from the global fit. As a point of comparison, we 
display the number of data in a histogram, for both $pp$ and $np$ 
scattering, in Fig.~\ref{fig:g4}, together with the number of measured 
observables. Clearly, the data constraints for a global fit are 
significantly reduced beyond 1.1~GeV, for $np$ scattering, and beyond 
about 2.5~GeV, for $pp$ scattering. In Fig.~\ref{fig:g5}, the $\chi^2$ 
difference between global and single-energy fits is also compared. This 
gives a qualitative picture of the fit stability which, as expected, 
improves with the number of data constraints. 
%%%%%%%%%%%%%%%%%%%%%%%%%%%%%%%%%%%%%%%%%%%%%%%%%%%
%%%%%%%%%%%%%%%%%%%%%%%%%%%%%%%%%%%%%%%%%%%%%%%%%%%%
-%%%%%%%%%%%%%%%%%%%%%%%%%%%%%%%%%%%%%%%%%%%%%%%%%%%%
%%%%%%%%%%%%%%%%%%%%%%%%%%%%%%%%%%%%%%%%%%%%%%%%%%%%

%%%%%%%%%%%%%%%%%%%%%%%%%%%%%%%%%%%%%%%%%%%%%%%%
\section{Amplitudes}
\label{sec:Ampl}

We are using the notation of Ref.~\cite{Bystricky98,Ball98} and write
the scattering matrix, $M(\vec k_f, \vec k_i)$, as
\begin{eqnarray*}
M(\vec k_f , \vec k_i) 
 & = & {1\over 2} [ (a + b) + (a - b)~(\vec \sigma_1 \cdot \vec n)
              ~(\vec \sigma_2 \cdot \vec n)  \nonumber \\
 & + & (c + d)~(\vec \sigma_1 \cdot \vec m )
              ~(\vec \sigma_2 \cdot \vec m ) \nonumber \\
 & + & (c - d)~(\vec \sigma_1 \cdot \vec l )
              ~(\vec \sigma_2 \cdot \vec l ) \nonumber \\
 & + &  e~(\vec \sigma_1 + \vec \sigma_2 )\cdot \vec n ] ,
        \label{dra1}
\end{eqnarray*}
where $\vec k_f$ and $ \vec k_i$ are the scattered and incident 
momenta in the c.m. system, and
\[ \vec n = { { \vec k_i \times \vec k_f }
        \over {|\vec k_i \times \vec k_f |} }
  \; , \;
   \vec l = { { \vec k_i + \vec k_f  }
        \over {|\vec k_i + \vec k_f | } }
  \; , \;
   \vec m = { { \vec k_f - \vec k_i  }
        \over {|\vec k_f - \vec k_i | } } .
 \]

Writing the scattering matrix in this form, any $pp$ observable
can be expressed in terms of the five complex amplitudes $a$
through $e$. If a sufficient number of independent observables
are measured precisely at a given energy and angle, these
amplitudes can be determined up to an overall undetermined
phase. The advantage of this method is its model independence;
nothing beyond the data is required to determine a solution.
In addition, once the amplitudes are found, any further
experimental quantity can be predicted at the energy-angle
points of the DAR. In practice, however, experimental uncertainties
allow for multiple solutions comparable in their representation of
the data. 

\begin{figure*}[th]
\centerline{
\includegraphics[height=0.75\textwidth, angle=90]{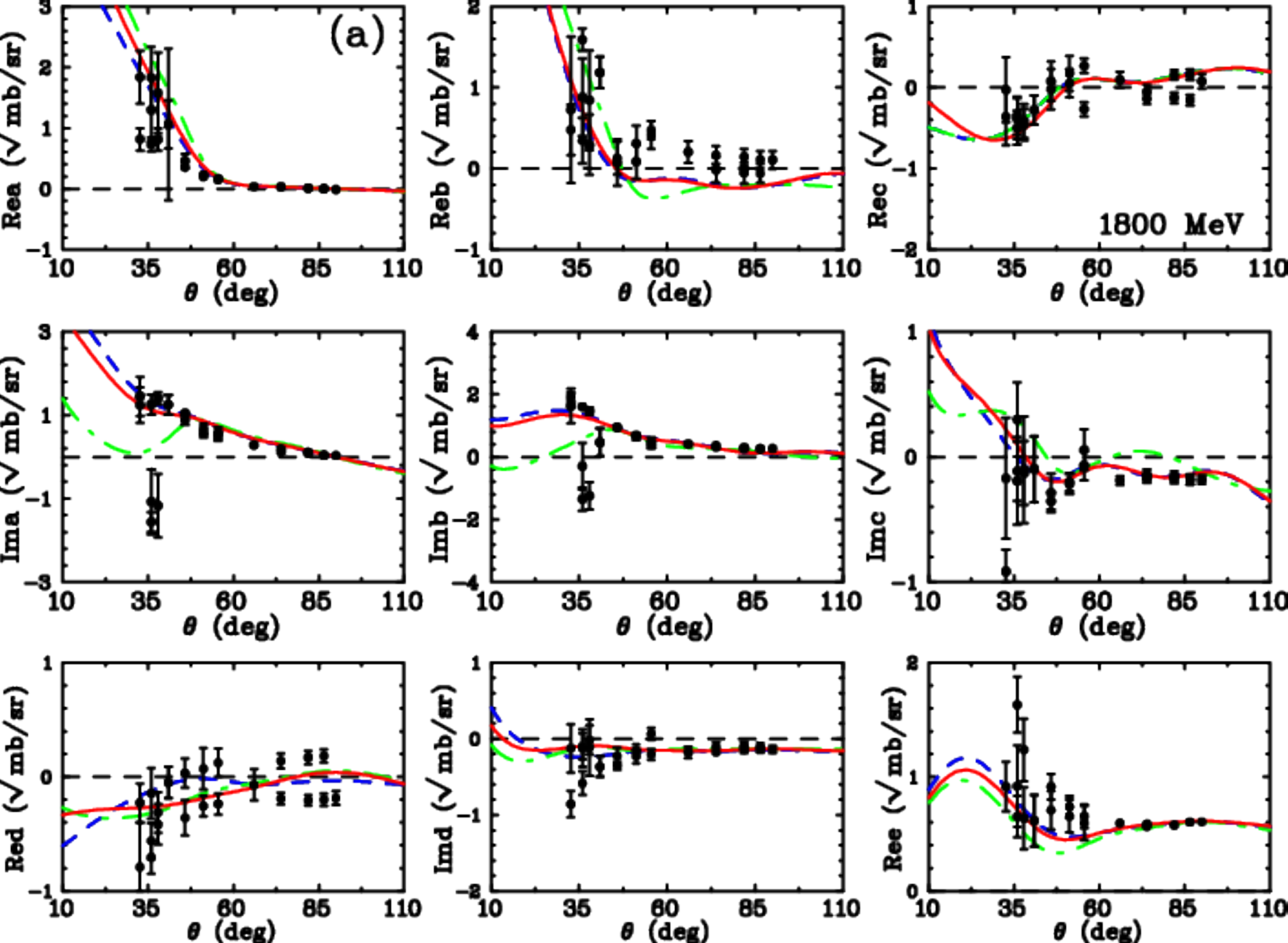}}
\centerline{
\includegraphics[height=0.75\textwidth, angle=90]{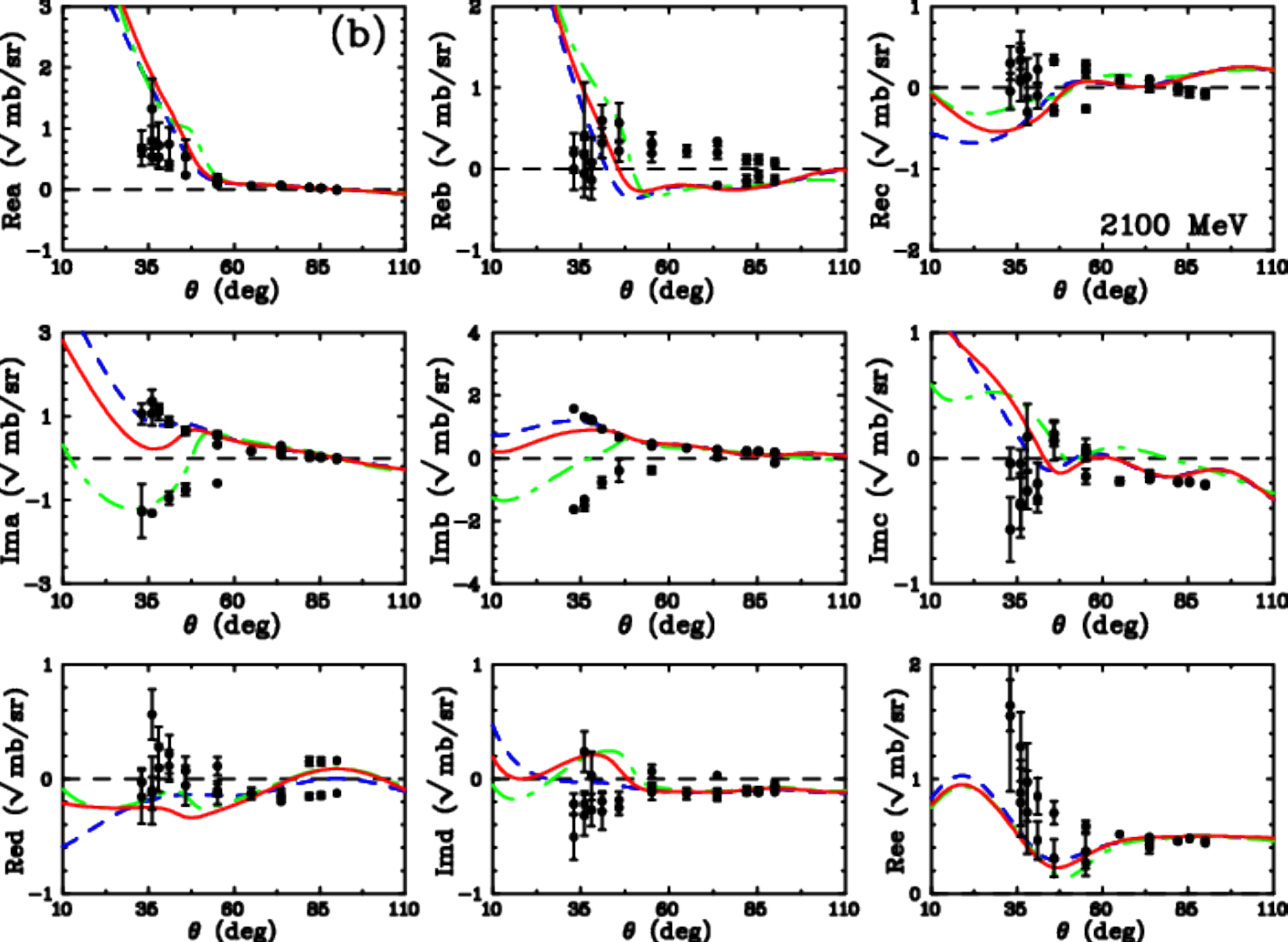}}
\vspace{10mm}
\caption{(Color online) DAR of
        $a$ to $e$ for $pp$ elastic scattering at (a)
        T$_{lab}$ = 1.8~GeV and (b) T$_{lab}$ = 2.1~GeV as
        a function of c.m. scattering angle.  Real and
        imaginary parts of the Saclay
        amplitudes~\protect\cite{Bystricky98} are shown as black
        filled circles. Notation for SAID energy-dependent
        amplitudes as in Fig.~\protect\ref{fig:g1}.
        \label{fig:g6}}
\end{figure*}

\begin{figure*}[th]
\centerline{
\includegraphics[height=0.75\textwidth, angle=90]{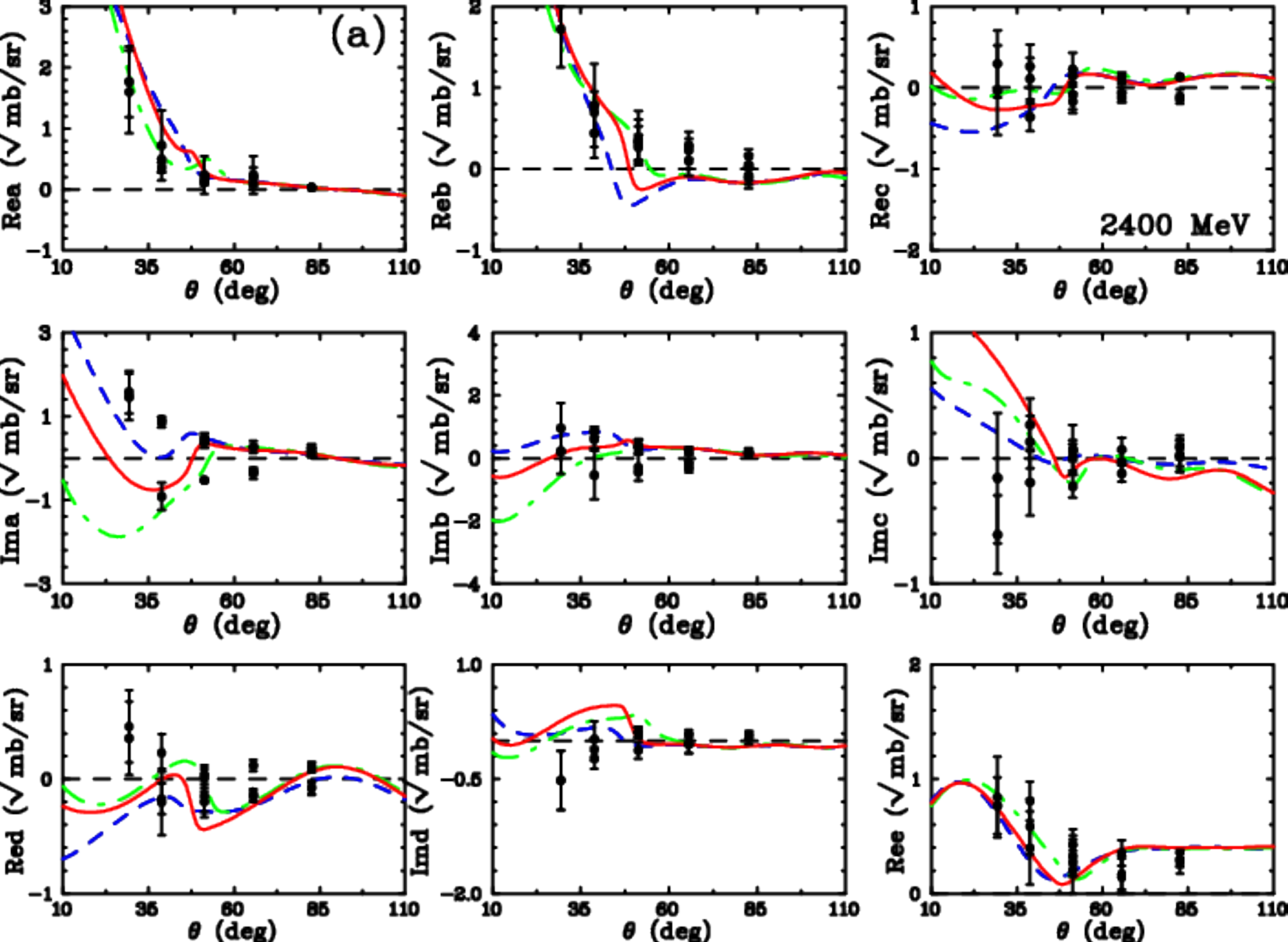}}
\centerline{
\includegraphics[height=0.75\textwidth, angle=90]{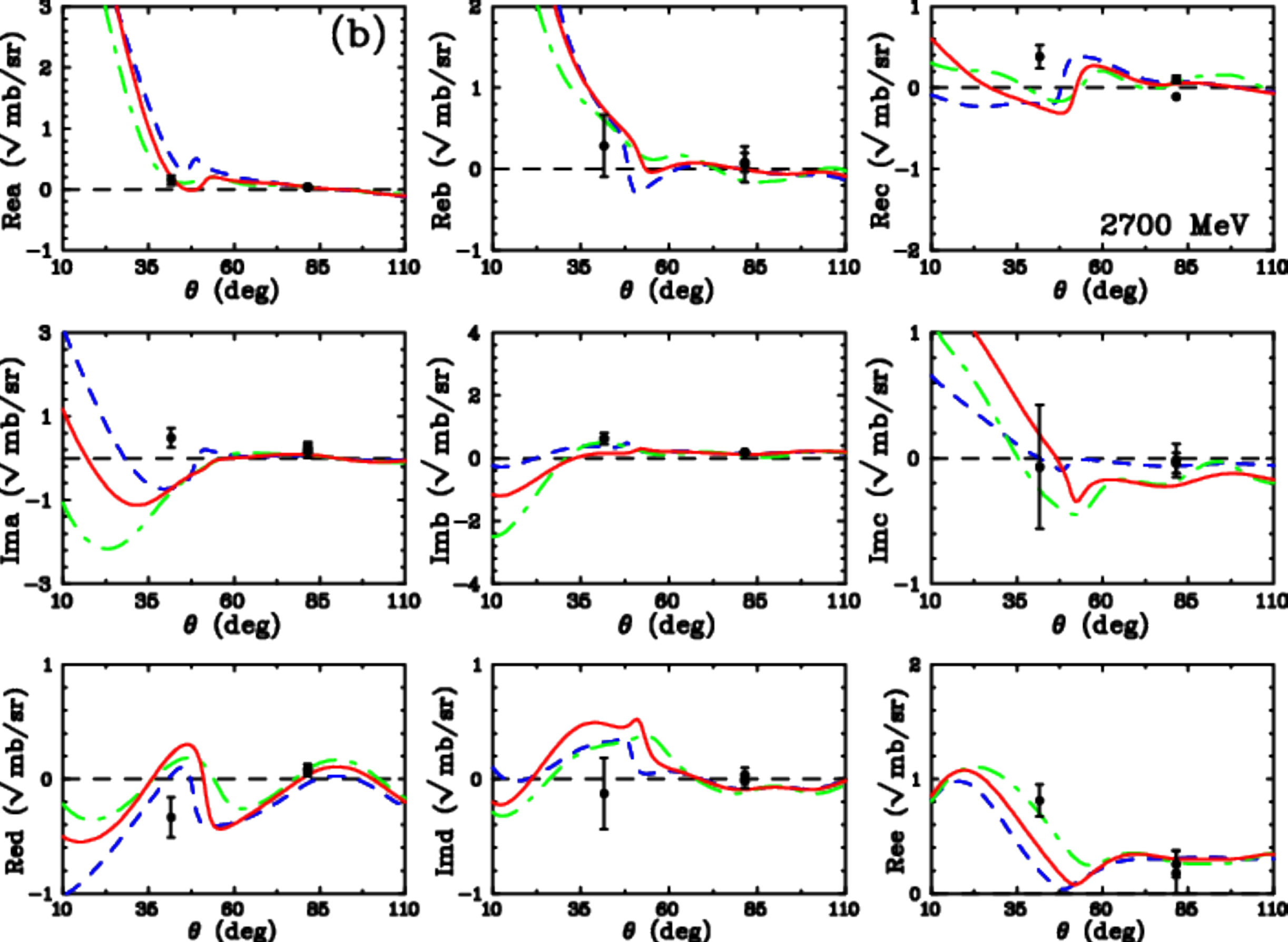}}
\vspace{10mm}
\caption{(Color online) DAR of
        $a$ to $e$ for $pp$ elastic scattering at (a) T$_{lab}$ 
	= 2.4~GeV and (b) T$_{lab}$ = 2.7~GeV as a function of 
	c.m. scattering angle. 
	Notation as in Fig.~\protect\ref{fig:g6}. \label{fig:g7}}
\end{figure*}

\begin{figure*}[th]
\centerline{
\includegraphics[height=0.75\textwidth, angle=90]{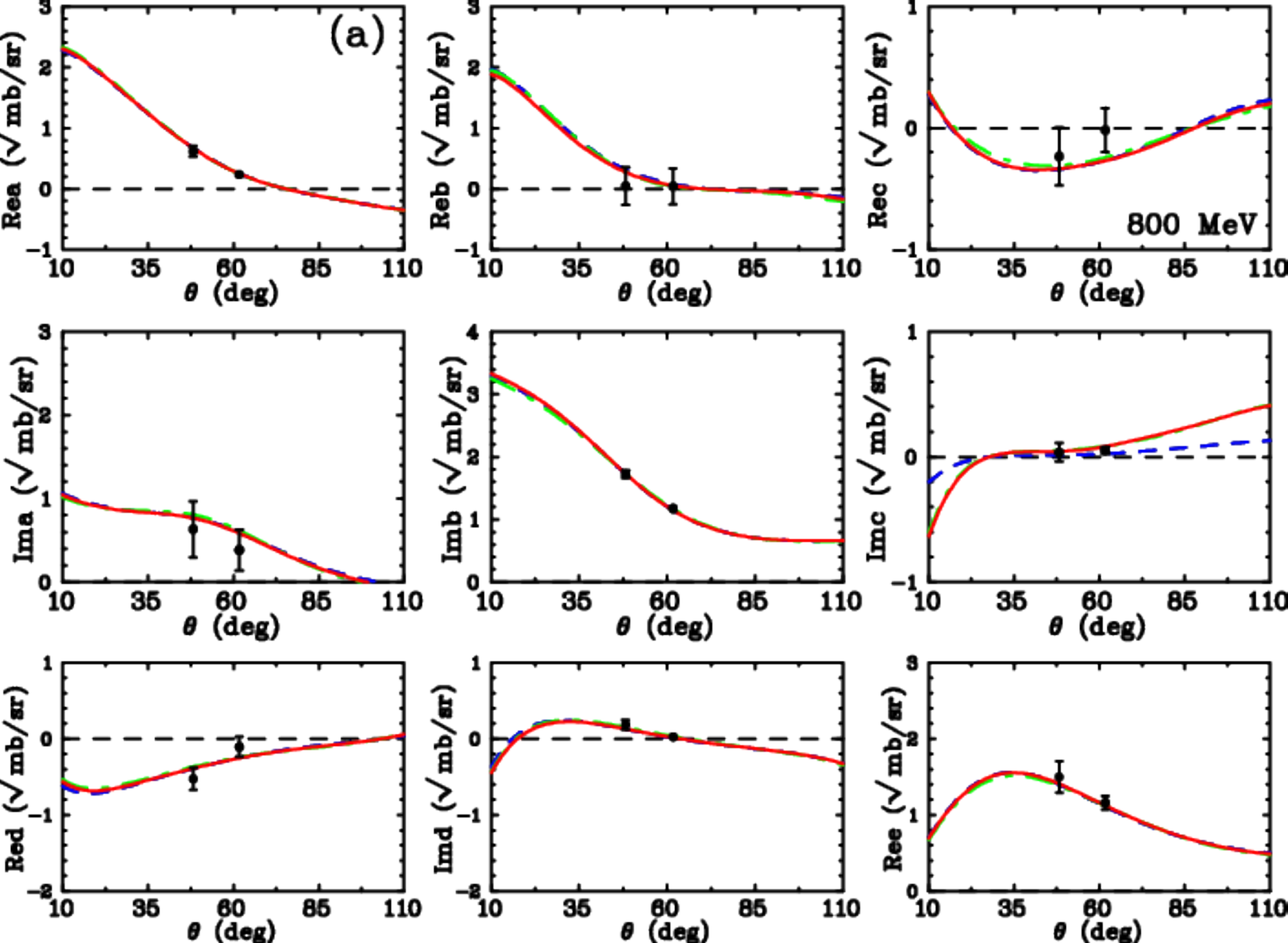}}
\centerline{
\includegraphics[height=0.75\textwidth, angle=90]{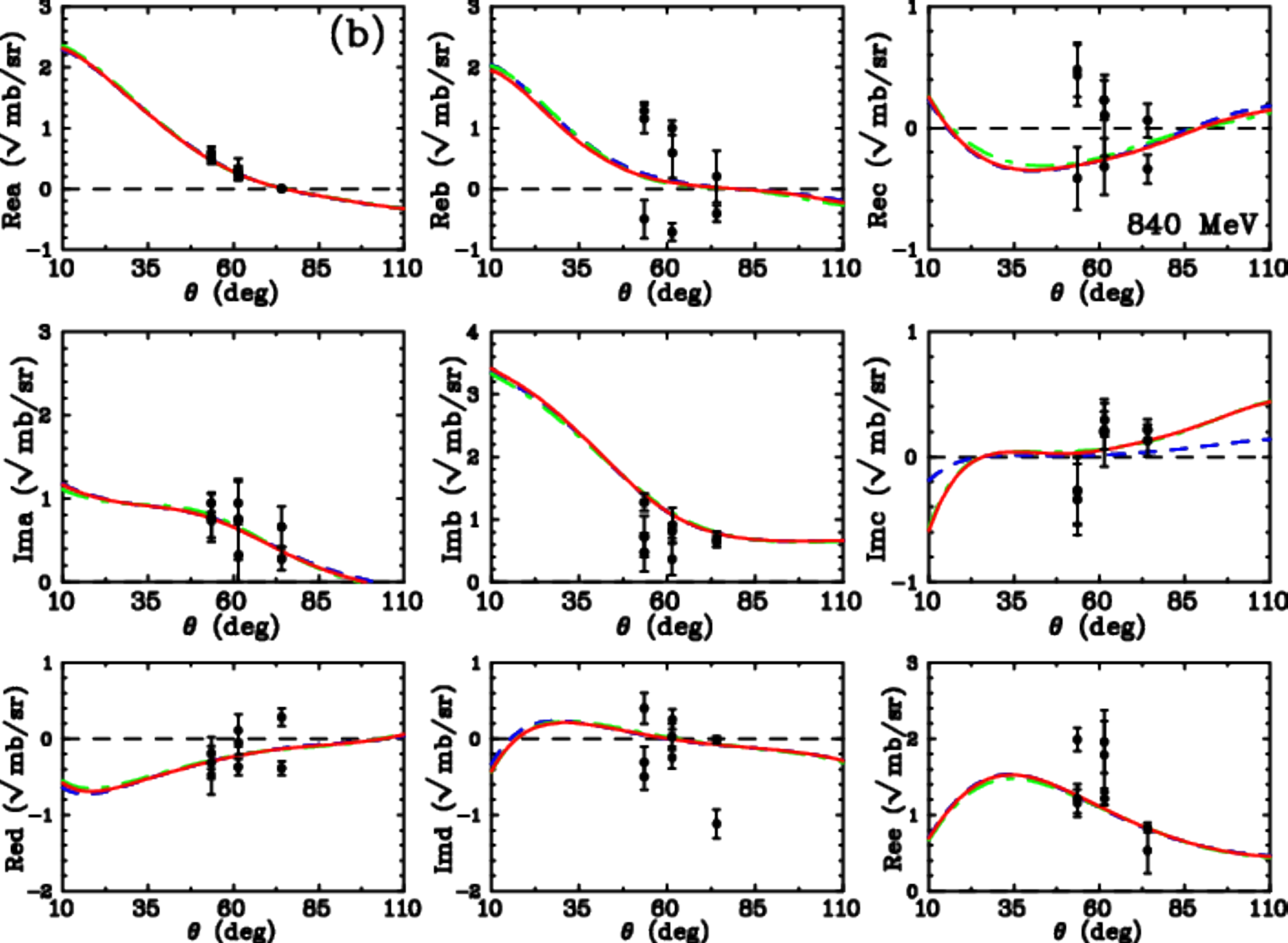}}
\vspace{10mm}
\caption{(Color online) DAR of
        $a$ to $e$ for $np$ elastic scattering at (a) T$_{lab}$ 
	= 0.80~GeV and (b) T$_{lab}$ = 0.84~GeV as a function 
	of c.m. scattering angle. Real and imaginary parts 
	of the Saclay amplitudes~\protect\cite{Ball98} are shown as black
        filled circles. Notation for SAID amplitudes as in
        Fig.~\protect\ref{fig:g1}. \label{fig:g8}}
\end{figure*}

\begin{figure*}[th]
\centerline{
\includegraphics[height=0.75\textwidth, angle=90]{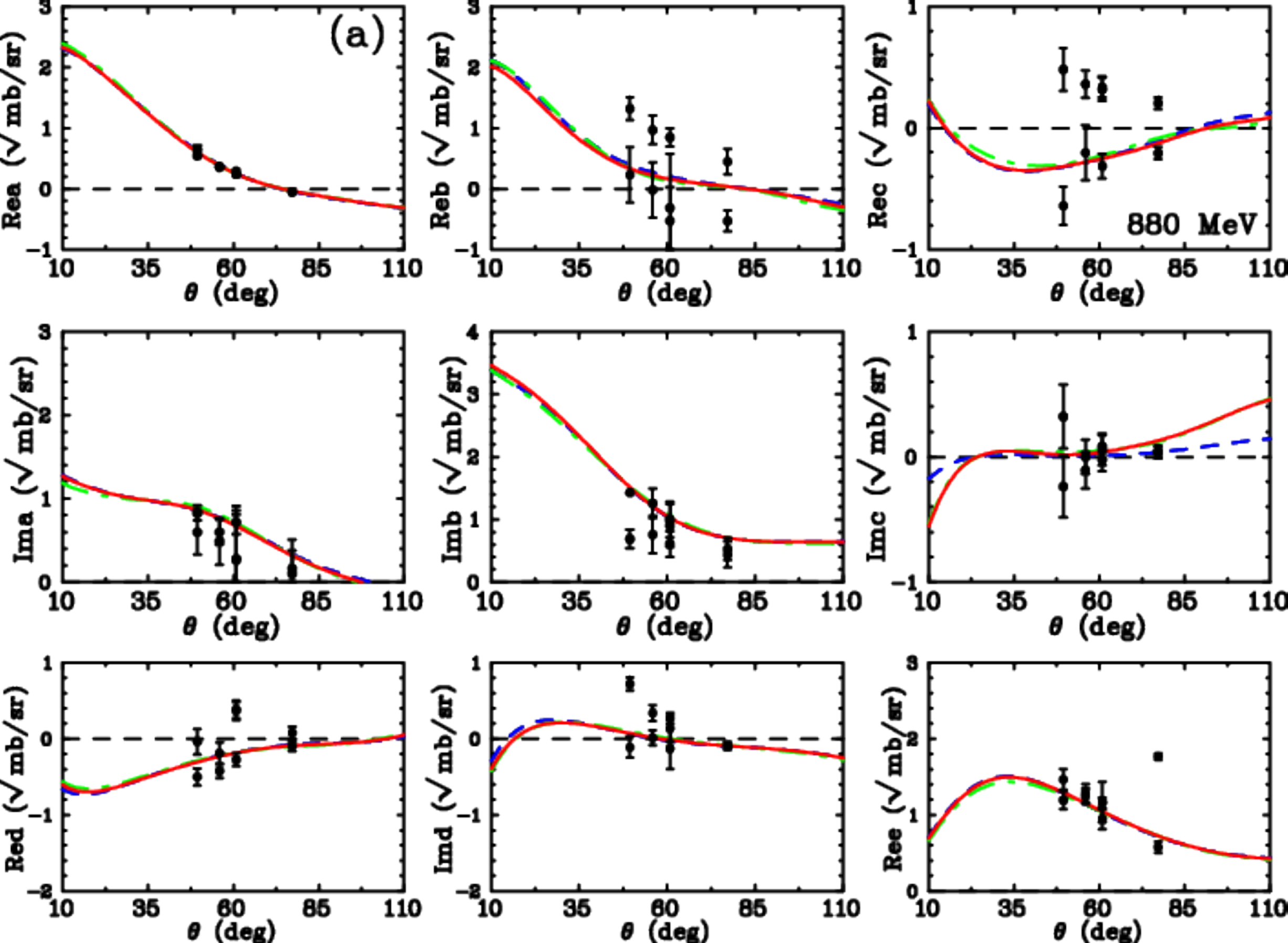}}
\centerline{
\includegraphics[height=0.75\textwidth, angle=90]{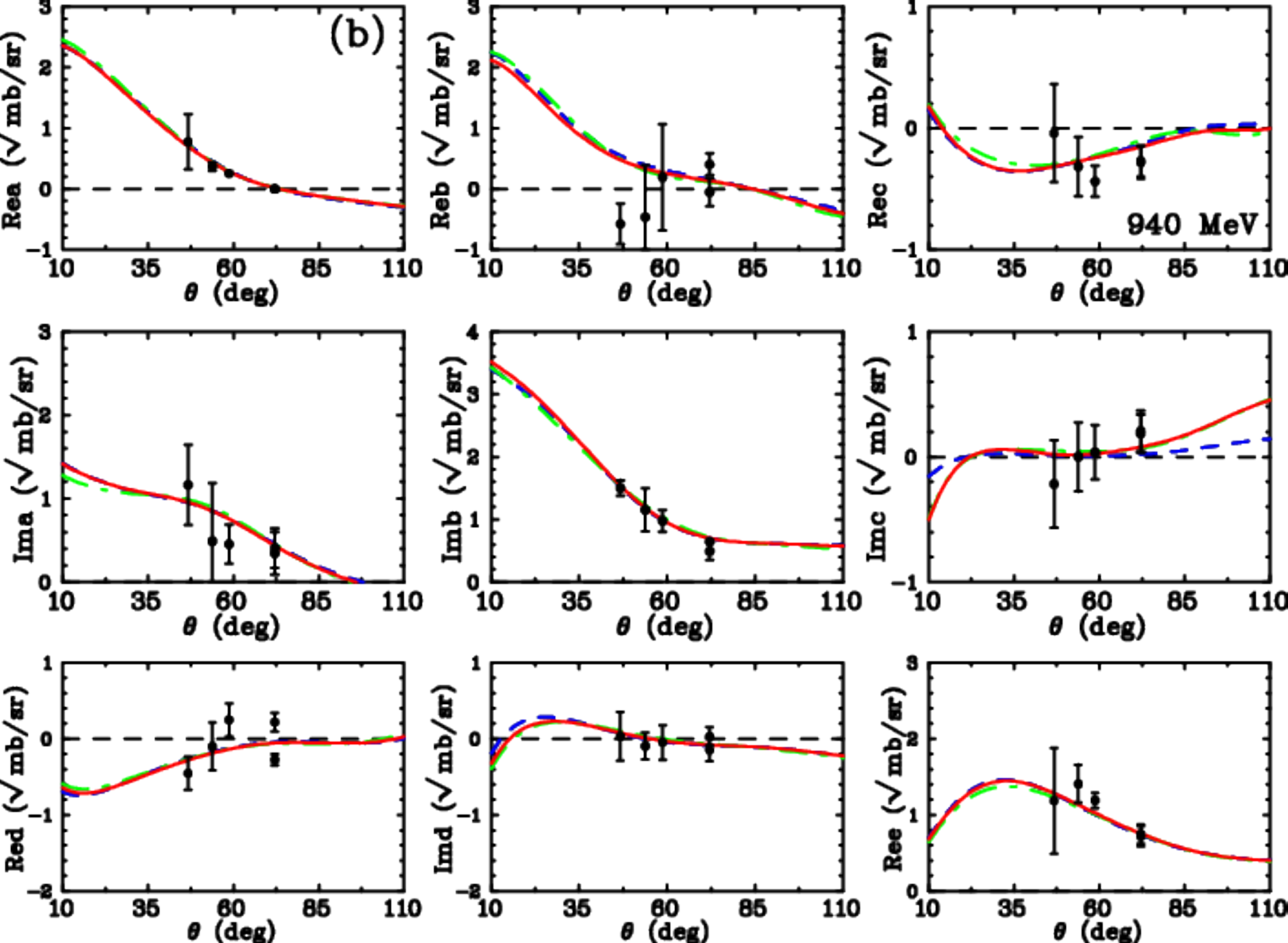}}
\vspace{10mm}
\caption{(Color online) DAR of
        $a$ to $e$ for $np$ elastic scattering at (a) T$_{lab}$ 
	= 0.88~GeV and (b) T$_{lab}$ = 0.94~GeV as a function 
	of c.m. scattering angle. Notation as in 
	Fig.~\protect\ref{fig:g8}. \label{fig:g9}}
\end{figure*}

\begin{figure*}[th]
\centerline{
\includegraphics[height=0.75\textwidth, angle=90]{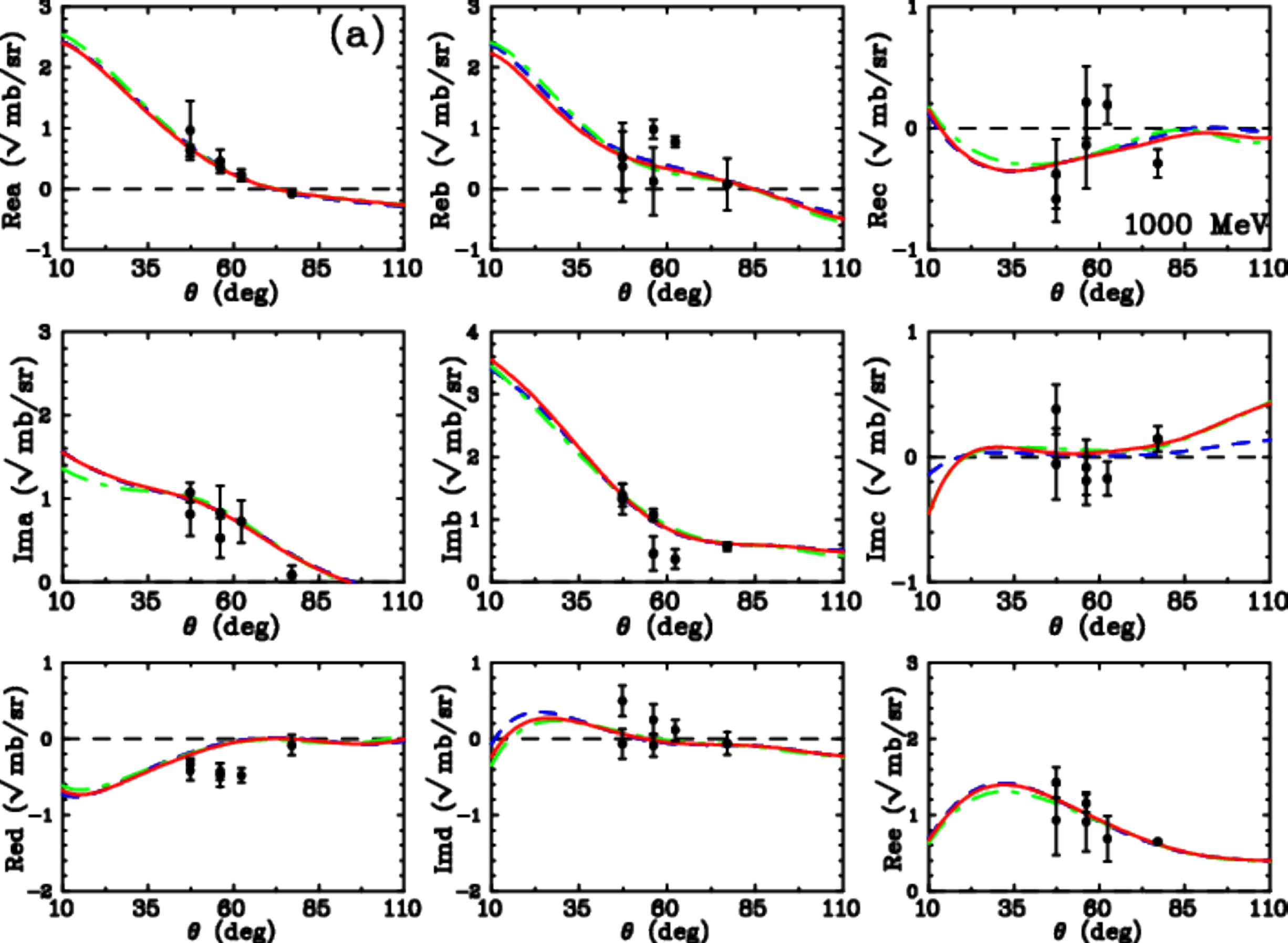}}
\centerline{
\includegraphics[height=0.75\textwidth, angle=90]{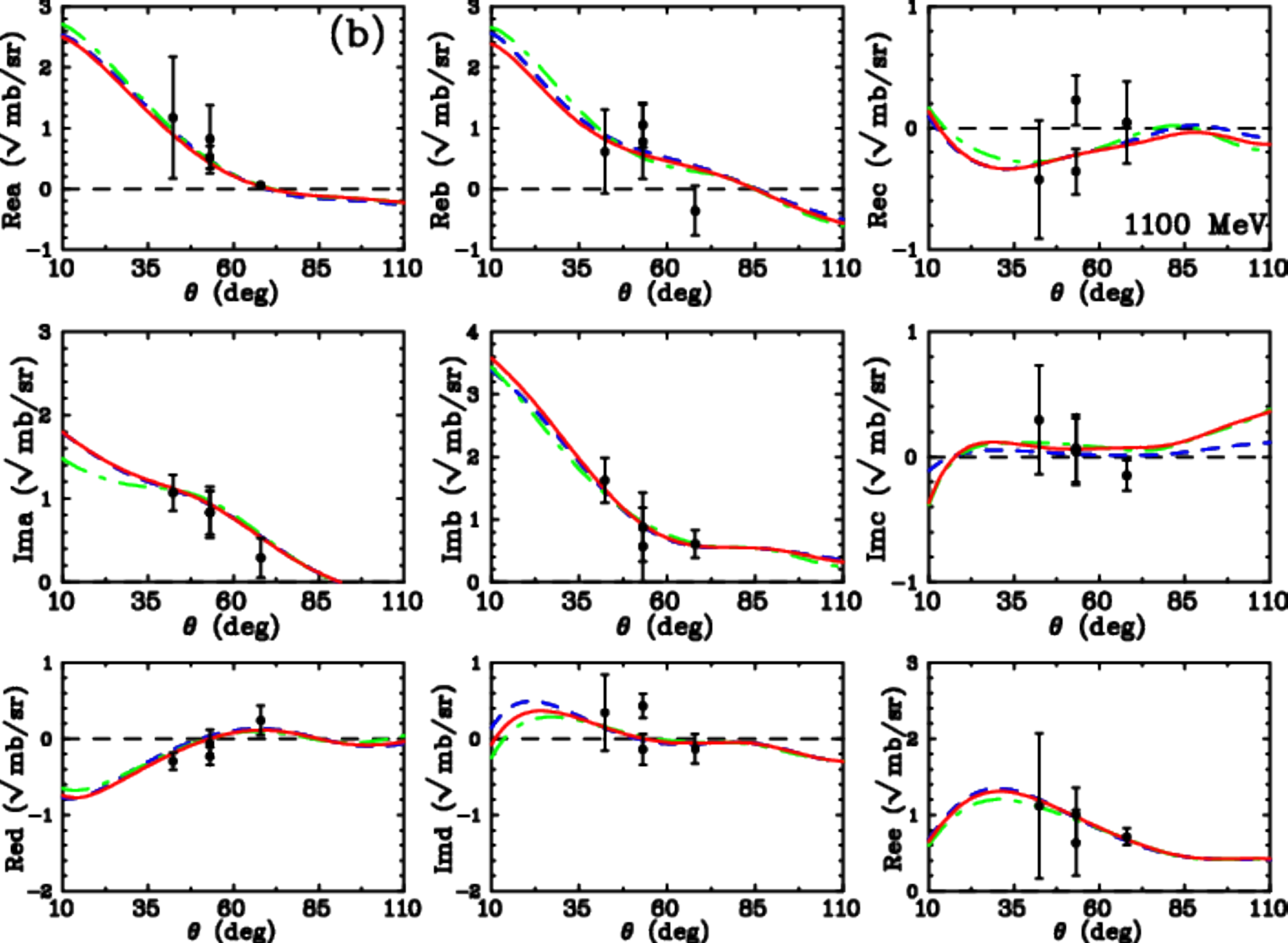}}
\vspace{10mm}
\caption{(Color online) DAR of
        $a$ to $e$ for $np$ elastic scattering at (a) 
	T$_{lab}$ = 1.00~GeV and (b) T$_{lab}$ = 1.10~GeV 
	as a function of c.m. scattering angle.
        Notation as in Fig.~\protect\ref{fig:g8}.
        \label{fig:g10}}
\end{figure*}

In Figs.~\ref{fig:g6} and \ref{fig:g7}, the $pp$ DAR amplitudes 
$a$ through $e$ of Ref.~\cite{Bystricky98} are plotted for energies 
between 1.8 and 2.7~GeV. These amplitudes have an overall phase 
ambiguity which is resolved, in the plots, by taking the $e$ amplitude 
to be real.
At the highest energy, with fewer data and observable types available,
results are only plotted at a pair of angles. At the other energies,
however, where multiple solutions form separate branches, the
fits SM16 and WF16 tend to branch as well. The DAR amplitudes of
Ref.~\cite{Ball98} for $np$ scattering, from 0.8 to 1.1~GeV,
show good overall agreement with the plotted fits. These are displayed
in Figs.~\ref{fig:g8}--\ref{fig:g10}.
%%%%%%%%%%%%%%%%%%%%%%%%%%%%%%%%%%%%%%%%%%%%%%%%%%%%
%%%%%%%%%%%%%%%%%%%%%%%%%%%%%%%%%%%%%%%%%%%%%%%%%%%%
%%%%%%%%%%%%%%%%%%%%%%%%%%%%%%%%%%%%%%%%%%%%%%%%%%%%
%%%%%%%%%%%%%%%%%%%%%%%%%%%%%%%%%%%%%%%%%%%%%%%%%%%%
%%%%%%%%%%%%%%%%%%%%%%%%%%%%%%%%%%%%%%%%%%%%%%%%%%%%
%%%%%%%%%%%%%%%%%%%%%%%%%%%%%%%%%%%%%%%%%%%%%%%%%%%%
%%%%%%%%%%%%%%%%%%%%%%%%%%%%%%%%%%%%%%%%%%%%%%%%%%%%
%%%%%%%%%%%%%%%%%%%%%%%%%%%%%%%%%%%%%%%%%%%%%%%%%%%%
%%%%%%%%%%%%%%%%%%%%%%%%%%%%%%%%%%%%%%%%%%%%%%%%%%%%

\begin{figure*}[th]
\centerline{
\includegraphics[height=0.45\textwidth, angle=90]{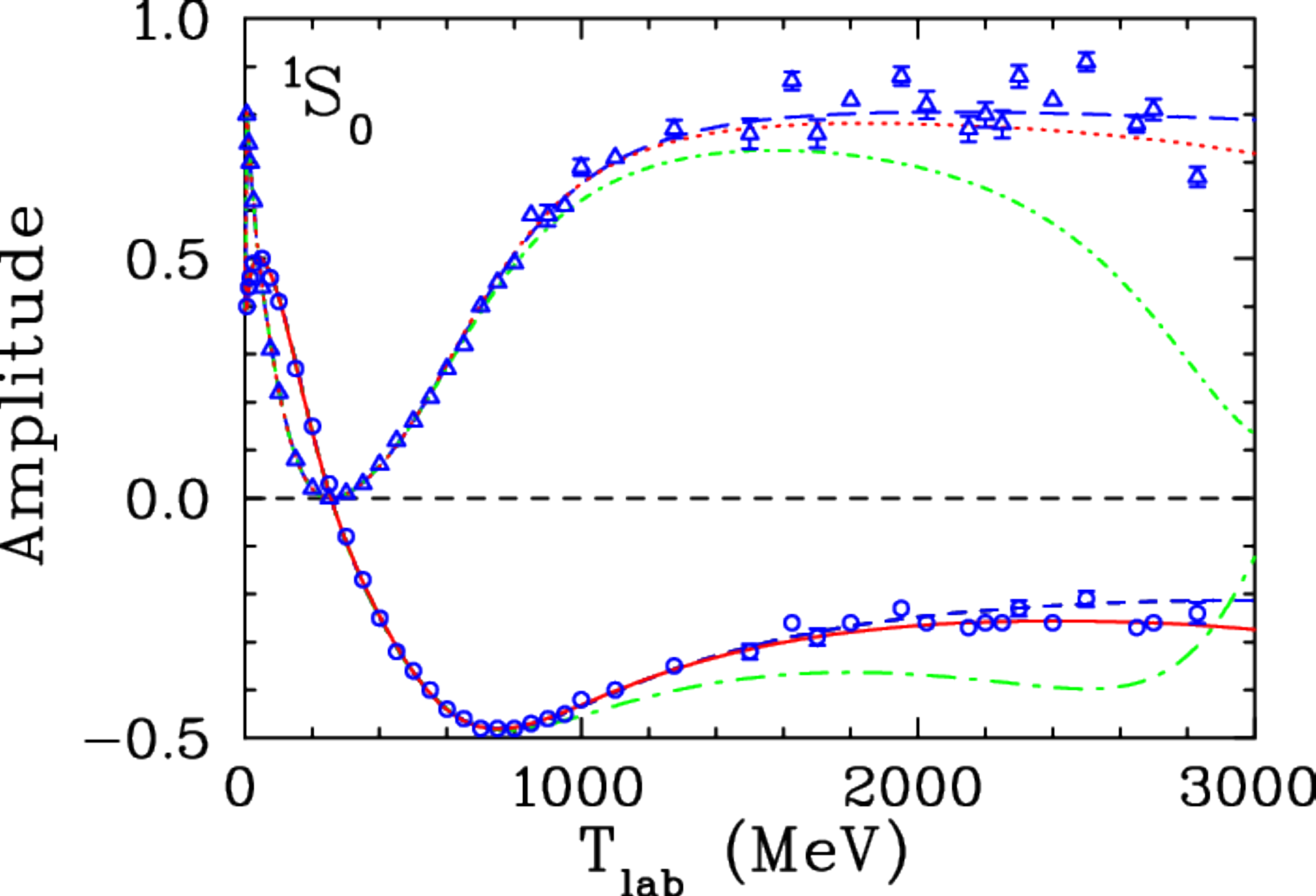}\hfill
\includegraphics[height=0.45\textwidth, angle=90]{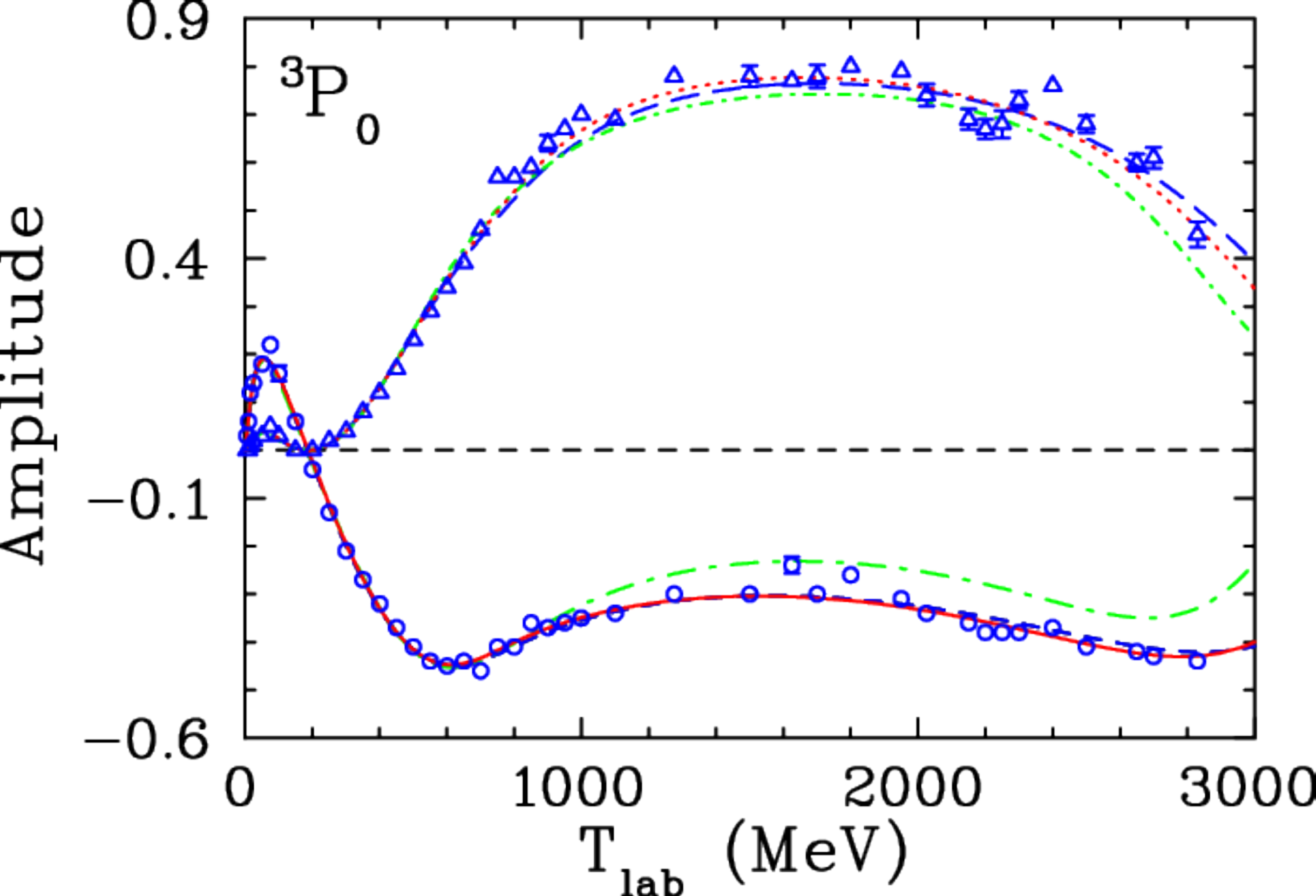}}
\centerline{
\includegraphics[height=0.45\textwidth, angle=90]{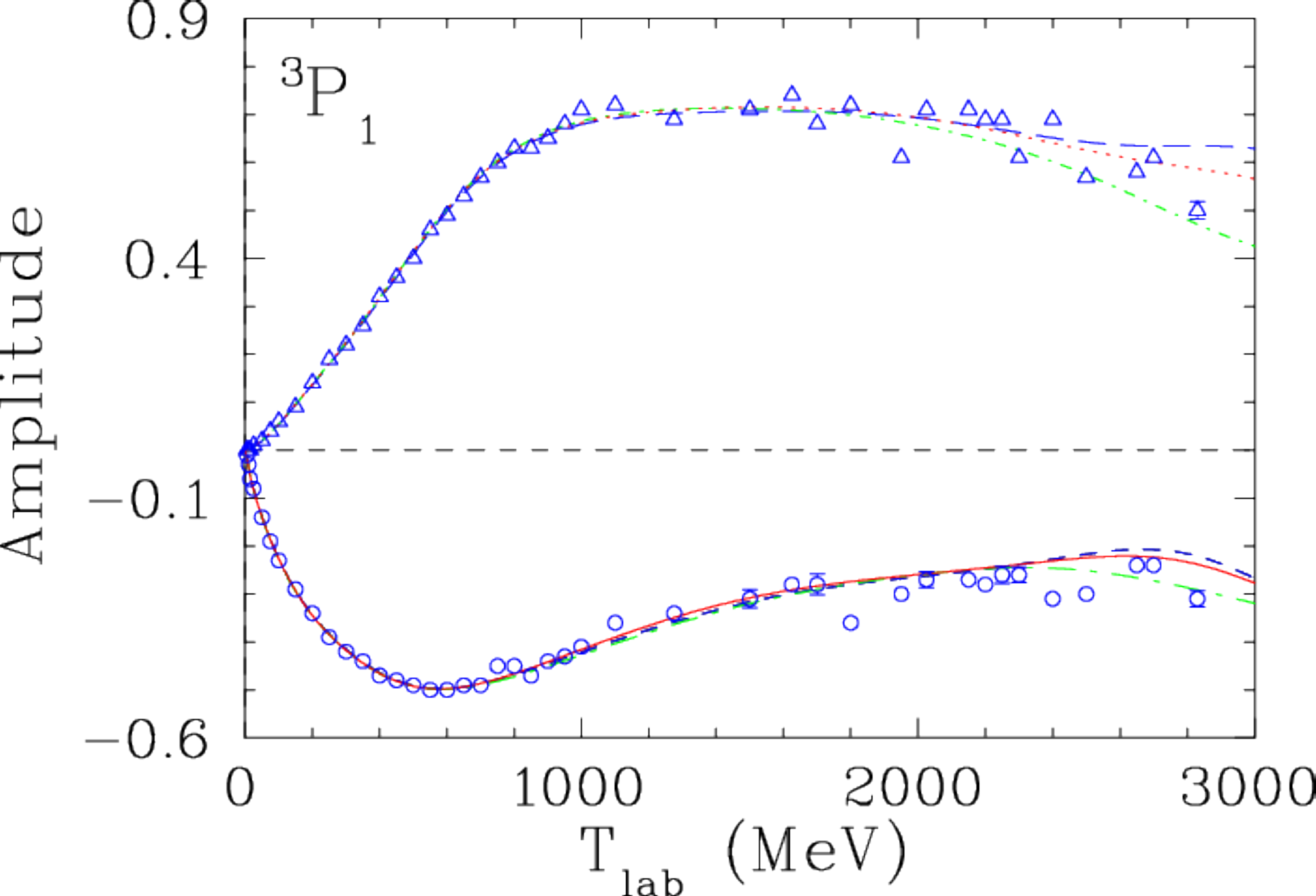}\hfill
\includegraphics[height=0.45\textwidth, angle=90]{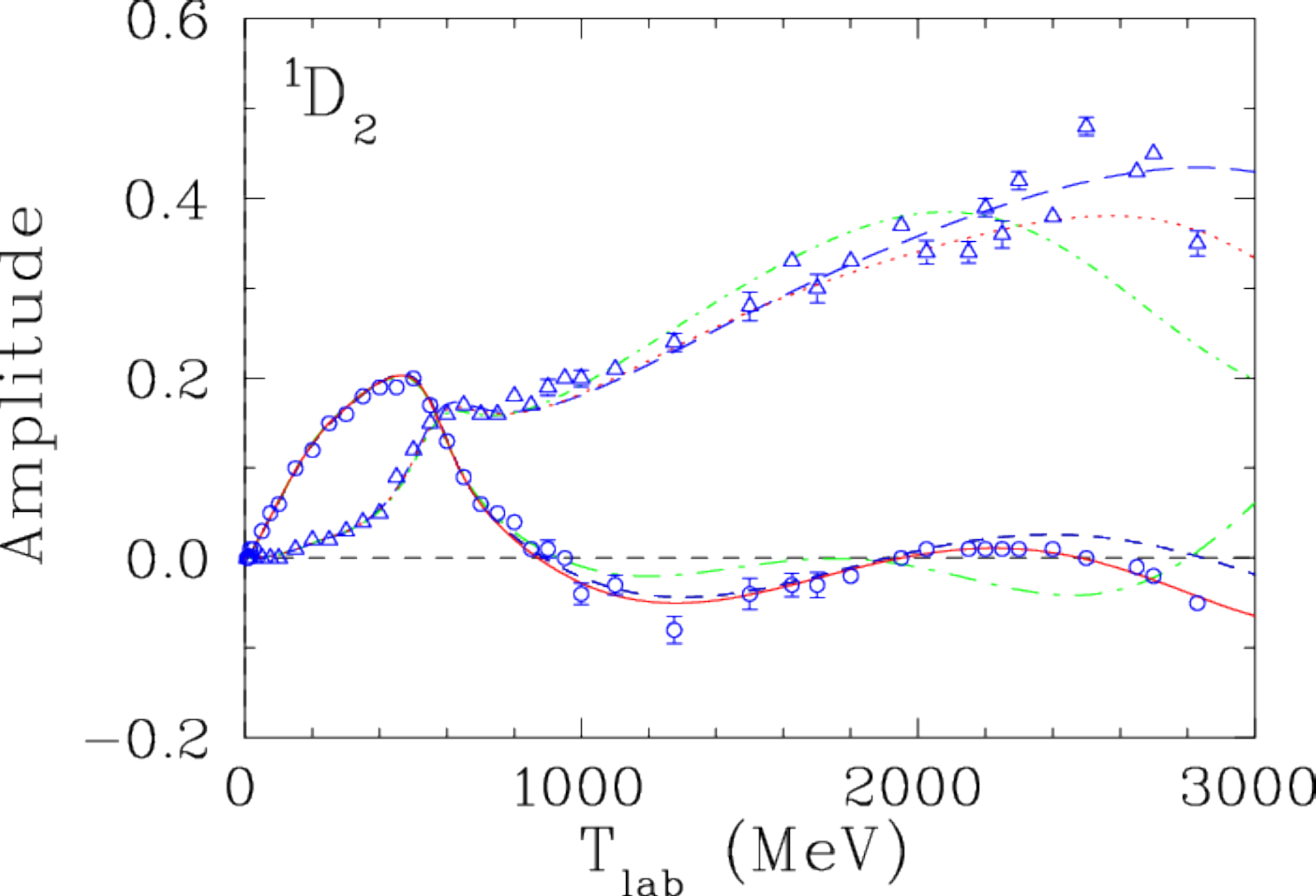}}
\centerline{
\includegraphics[height=0.45\textwidth, angle=90]{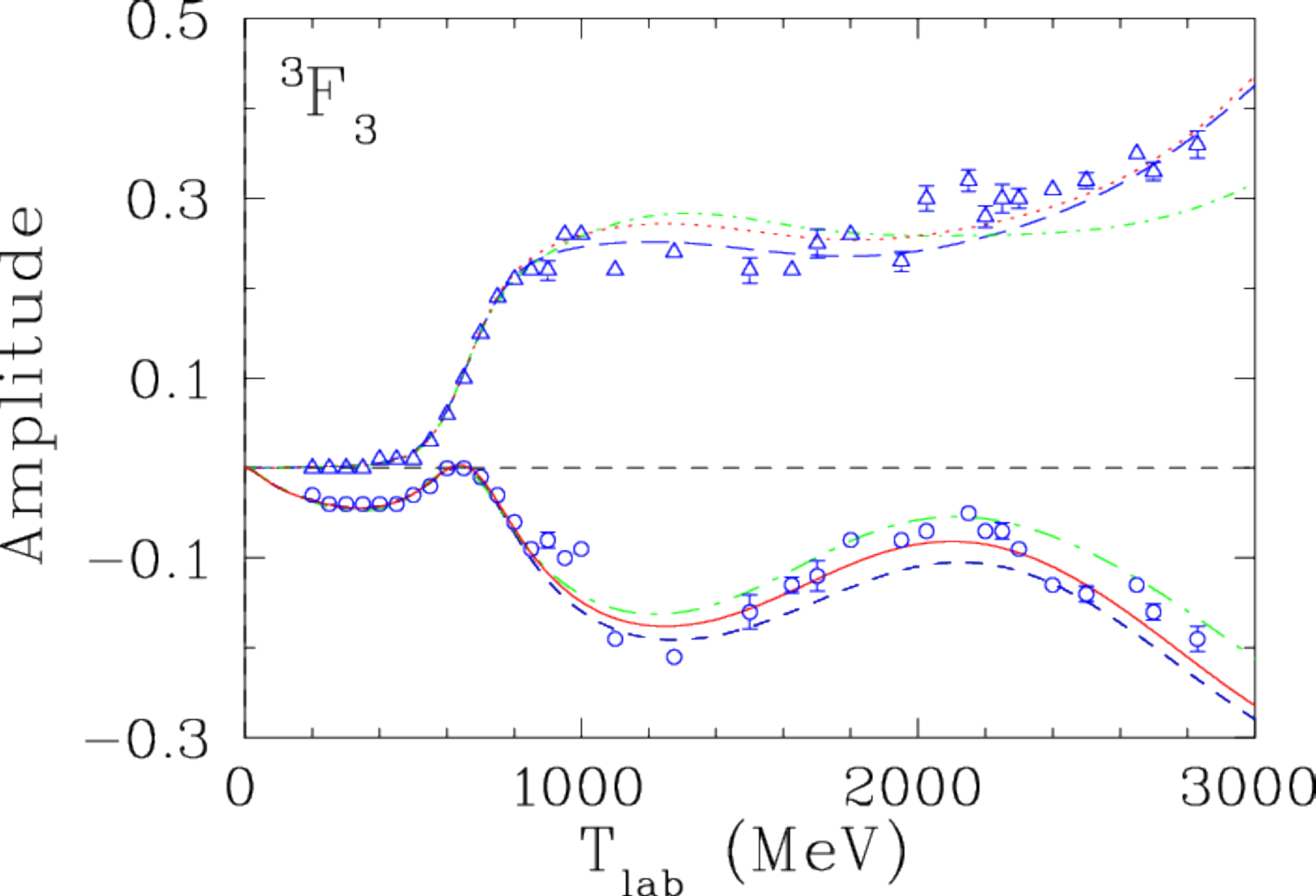}\hfill
\includegraphics[height=0.45\textwidth, angle=90]{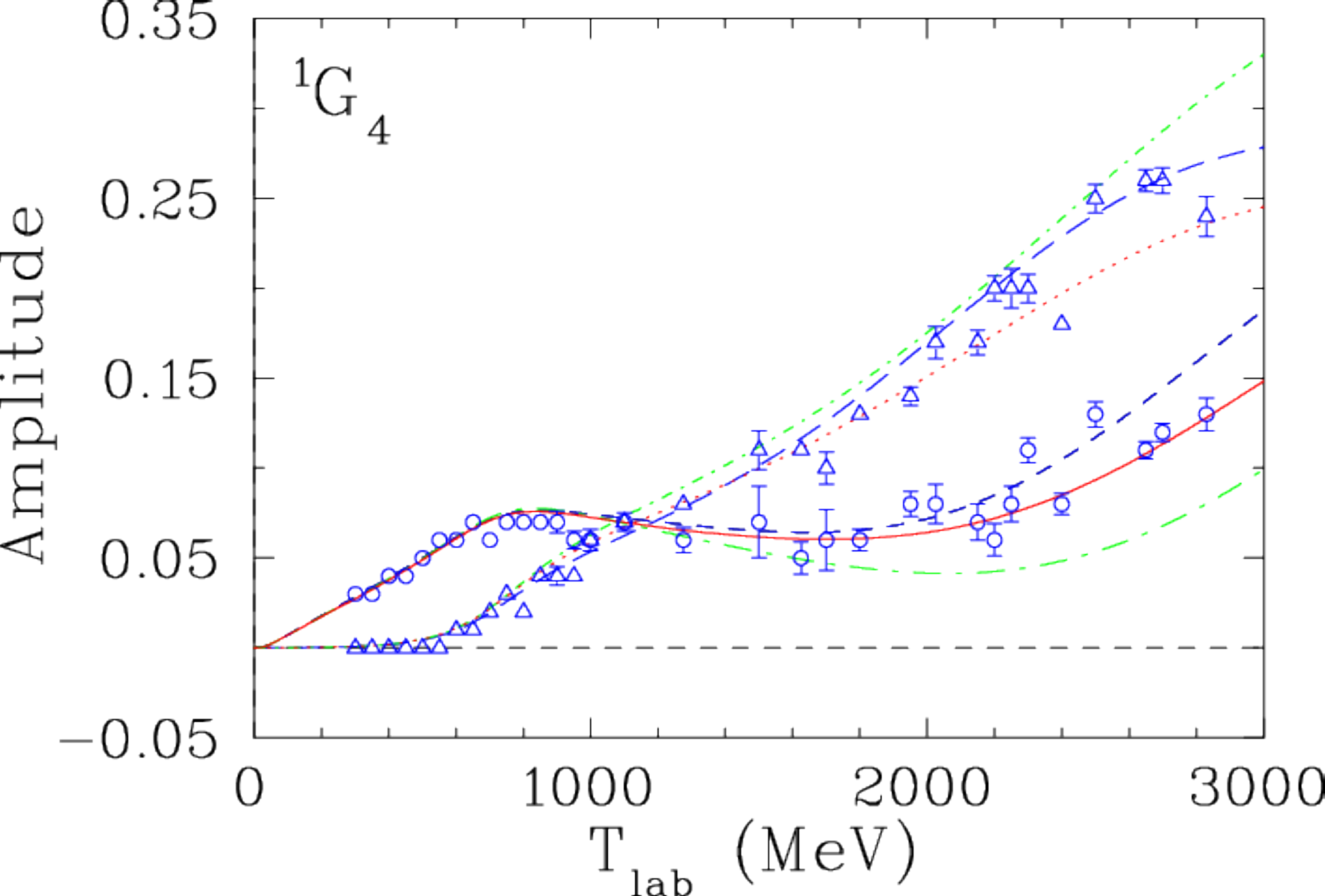}}
\centerline{
\includegraphics[height=0.45\textwidth, angle=90]{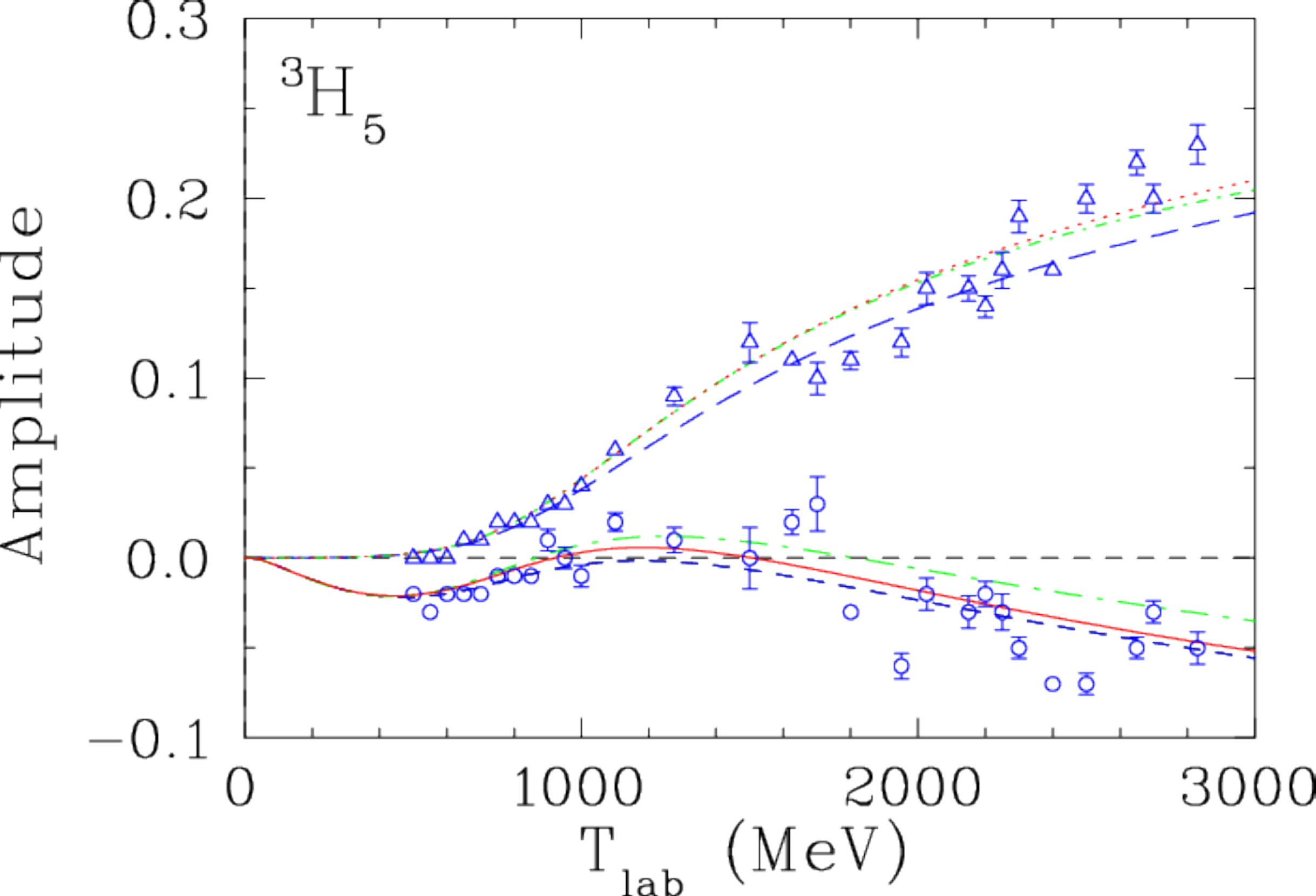}\hfill
}
\vspace{3mm}
\caption{(Color online) Dominant isovector partial-wave
        amplitudes from threshold to $T_{lab}$ = 3~GeV.
	Real (imaginary) parts of energy dependent solutions
        SM16, WF16, and SP07 are plotted as red solid (dotted),
        green long dot-dashed (short dot-dashed), and 
        blue short-dashed (long-dashed) lines, respectively. 
	Real (imaginary) parts of
	SES, corresponding to energy dependent fit SM16, are 
	plotted with blue open circles (triangles).
        All amplitudes are dimensionless. \label{fig:g11}}
\end{figure*}

\begin{figure*}[th]
\centerline{
\includegraphics[height=0.45\textwidth, angle=90]{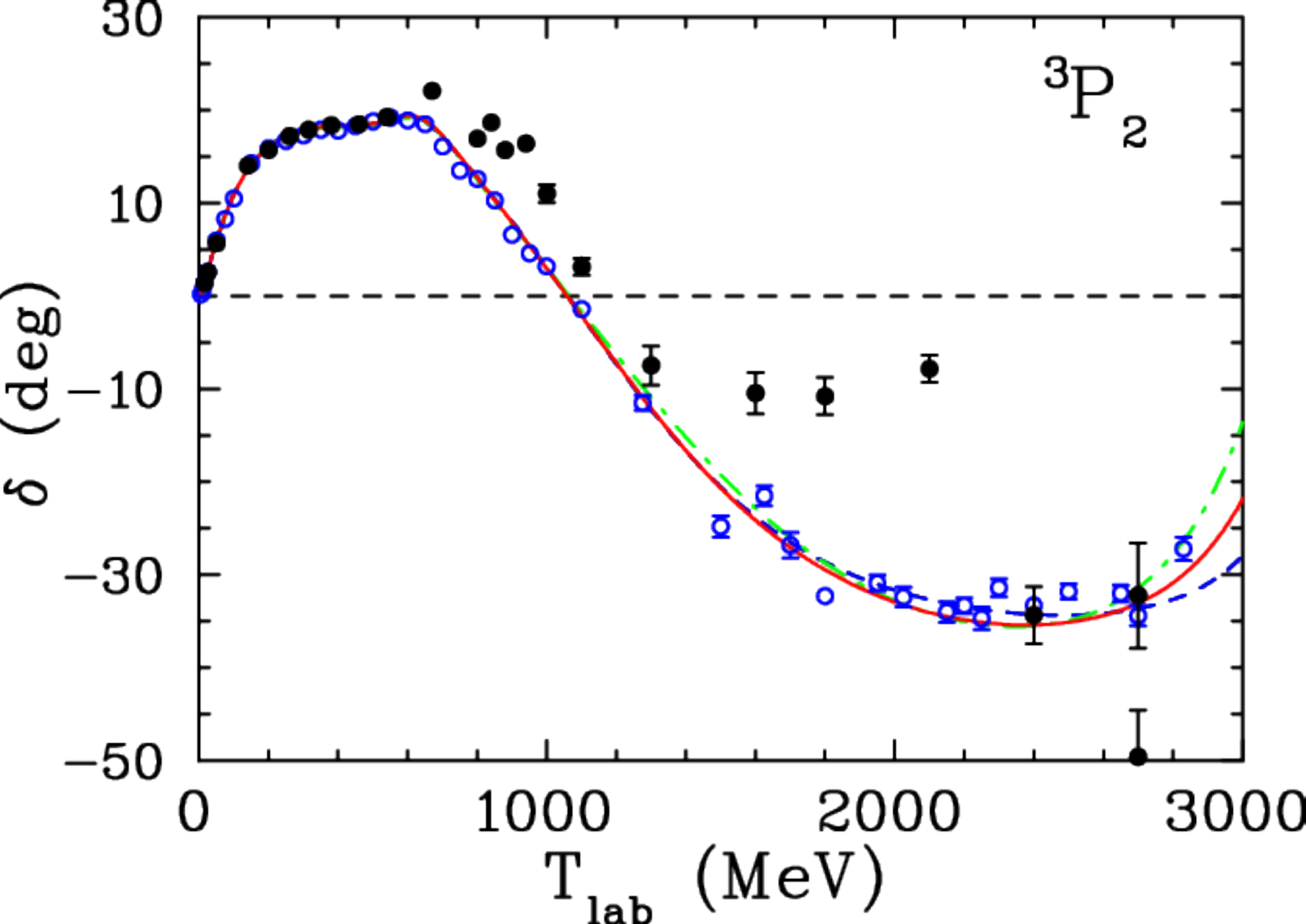}\hfill
\includegraphics[height=0.45\textwidth, angle=90]{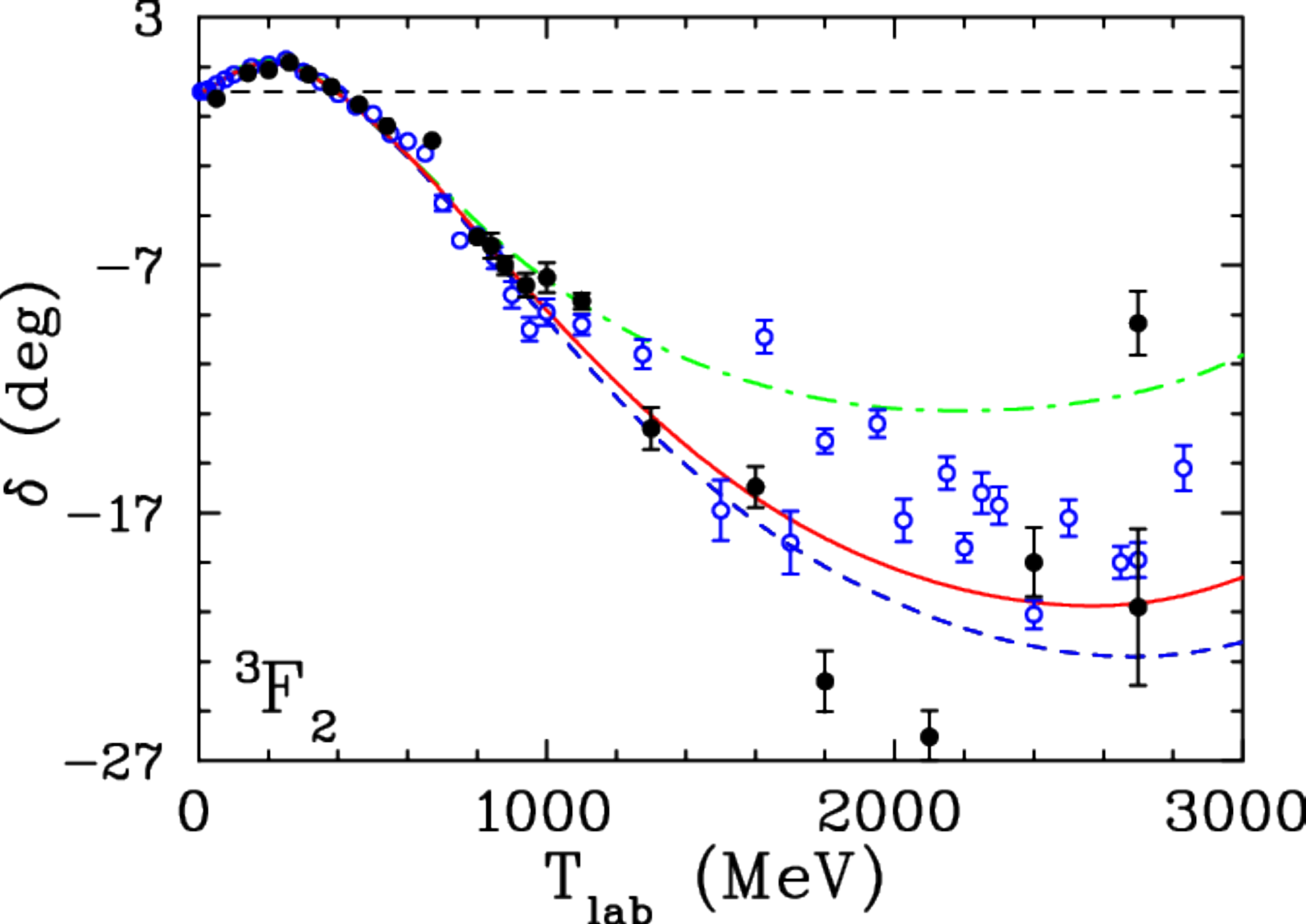}}
\centerline{
\includegraphics[height=0.45\textwidth, angle=90]{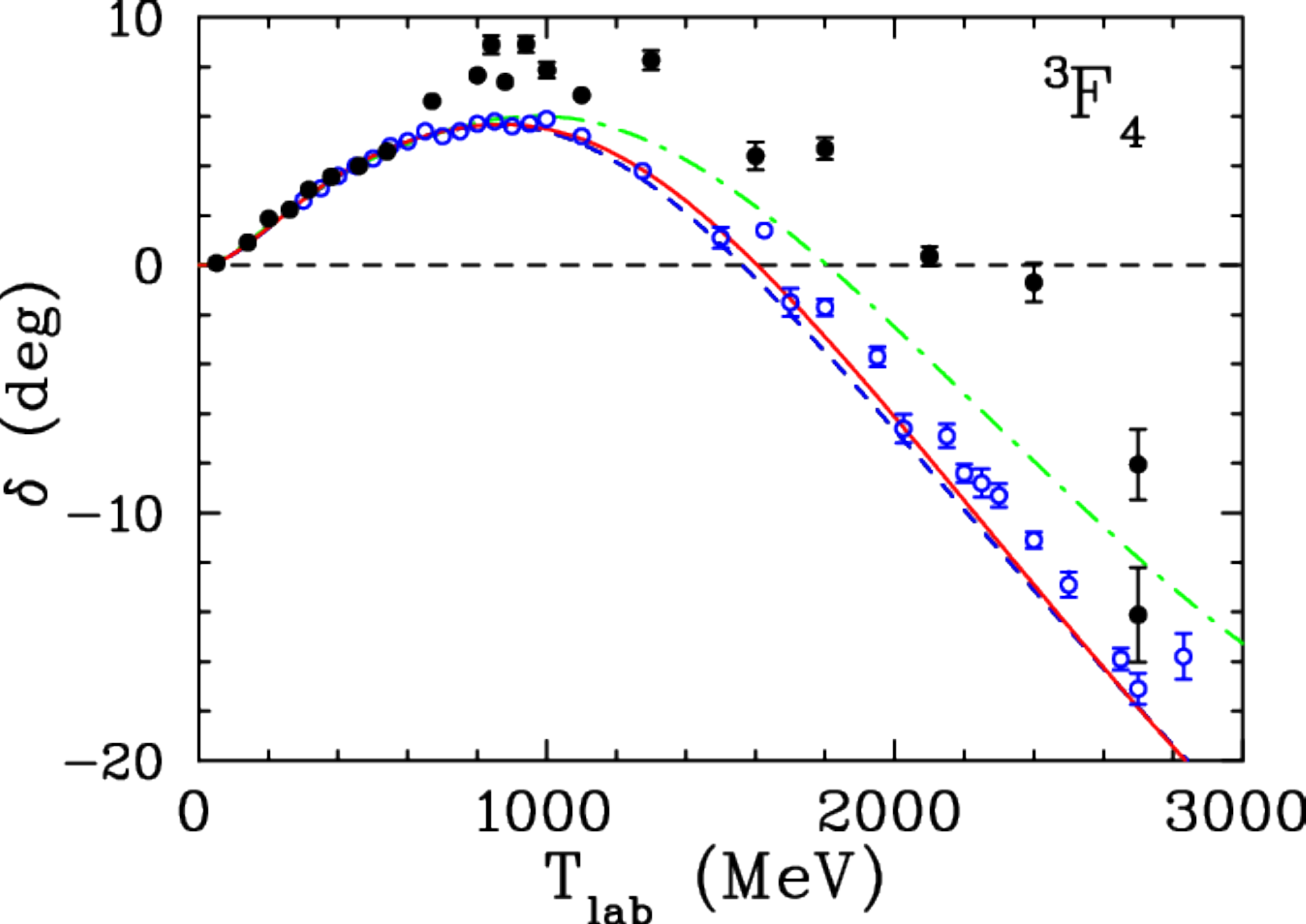}\hfill
\includegraphics[height=0.45\textwidth, angle=90]{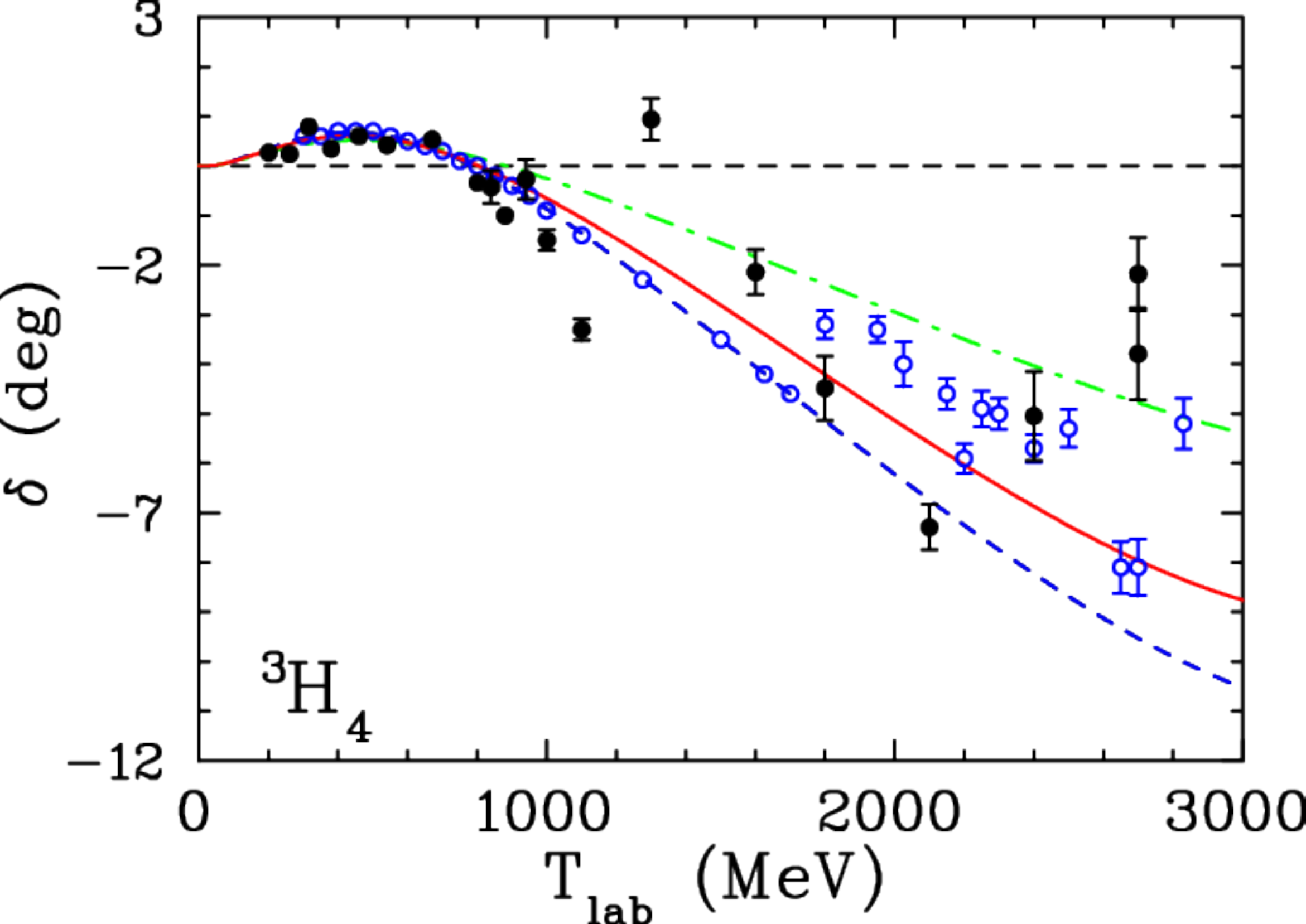}}
%\centerline{
%\includegraphics[height=0.45\textwidth, angle=90]{./fig/cf4l.eps}\hfill
%\includegraphics[height=0.45\textwidth, angle=90]{./fig/cf4m.eps}}
\vspace{3mm}
\caption{(Color online) Phase-shift parameters for dominant
        isovector partial-wave amplitudes from threshold to
        $T_{lab}$ = 3~GeV. Saclay SES
	phase-shifts~\protect\cite{Bystricky98} are shown 
	as black filled circles; SAID SES are plotted as 
        open blue circles. Notation for SAID energy-dependent
        amplitudes as in Fig.~\protect\ref{fig:g1}. \label{fig:g12}}
\end{figure*}

In Figs.~\ref{fig:g11} and \ref{fig:g12}, the dominant
isovector PWA results are plotted. Here we compare
the previous SP07 result to the present SM16 and
WF16 fits. Unlike the DAR results, a PWA is not strictly
model-independent and variations between fits are expected
with the onset of inelasticity and an increasing angular
momentum cutoff. However, for values of ${\rm T_{Lab}}$ 
below 1 GeV, fits have remained very stable. The largest
deviations are generally found in smaller partial waves and at the
highest fitted energies. Results for these and previous fits
may be found on the SAID website, together with the
associated database~\cite{SAID}.

In Fig.~\ref{fig:g11}, the dimensionless T-matrices are plotted
for the global fits and the SES results. The fits SM16 and SP07
are qualitatively similar and fall within the scatter of the
SES values. The weighted fit, WF16, constrained to better reproduce
the recent measurements, shows significant deviations from SM16
in the ${\rm ^1S_0}$ and ${\rm ^1D_2}$ partial waves. This effect
was noted in Ref.~\cite{mc16}. The behavior of the ${\rm ^1S_0}$ 
wave is remarkably similar to the much older fit, SM97, completed
in 1997. As displayed in Fig.~\ref{fig:g3}, SM97 
followed the polarized and unpolarized total cross sections more
closely than SP07. Thus, we find the ${\rm ^1S_0}$ 
wave to be particularly sensitive
to constraints imposed to fit the total cross sections and the 
forward differential cross sections.

In Fig.~\ref{fig:g12}, selected phase shifts are plotted for both the
global and SES fits and are compared to results from the 
Saclay SES~\cite{Bystricky98}. All determinations agree quite well
below 500 MeV and begin to scatter significantly above 1 GeV. A 
significant deviation between the SM16 and WF16 fits is visible in
the ${\rm ^3F_2}$ and this was noted, in Ref.~\cite{ba14}, to be the
result of constraints imposed by the recent forward $A_y$ 
measurements~\cite{ba14}.

\begin{figure*}[th]
\centerline{
\includegraphics[height=0.45\textwidth, angle=90]{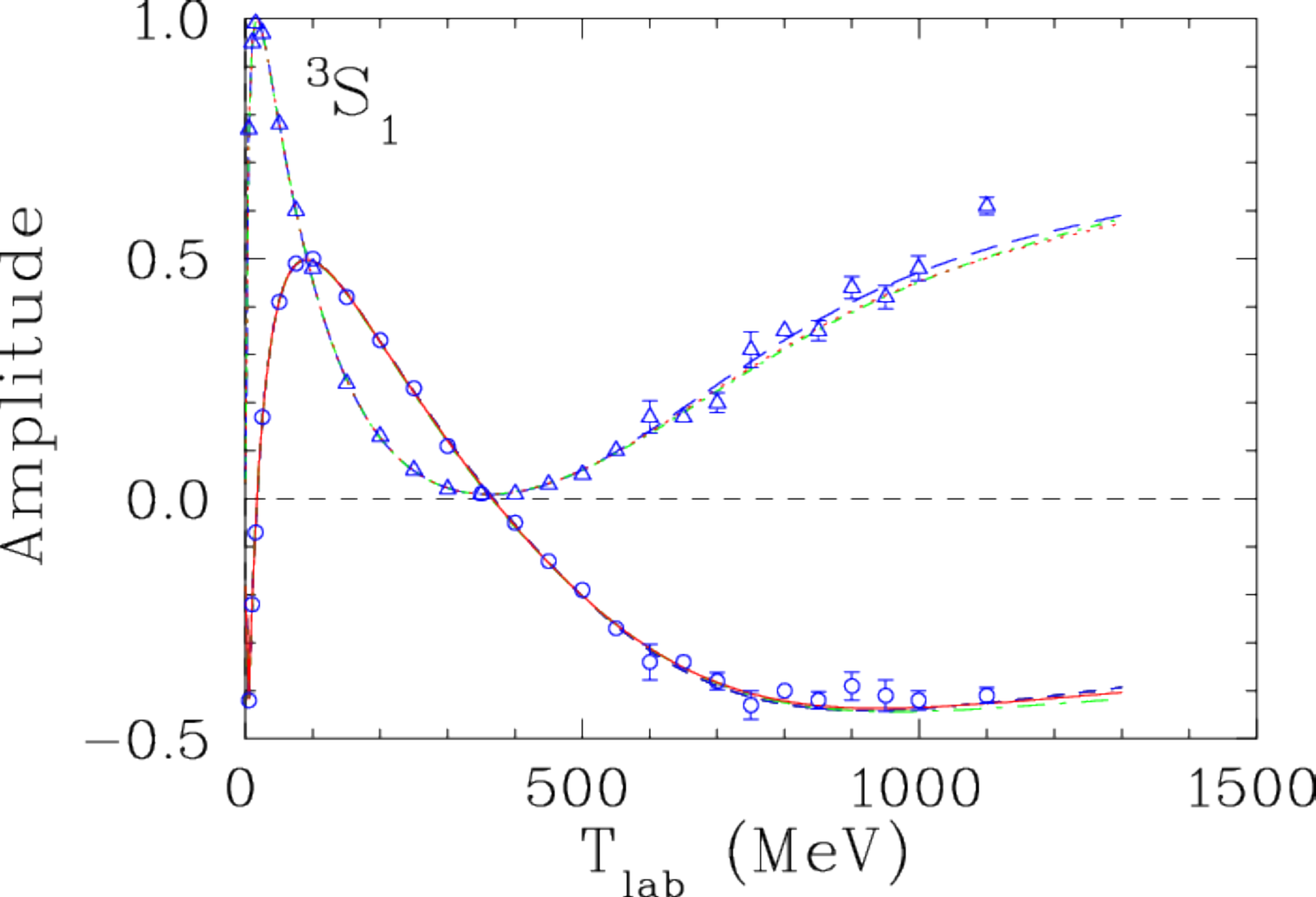}\hfill
\includegraphics[height=0.45\textwidth, angle=90]{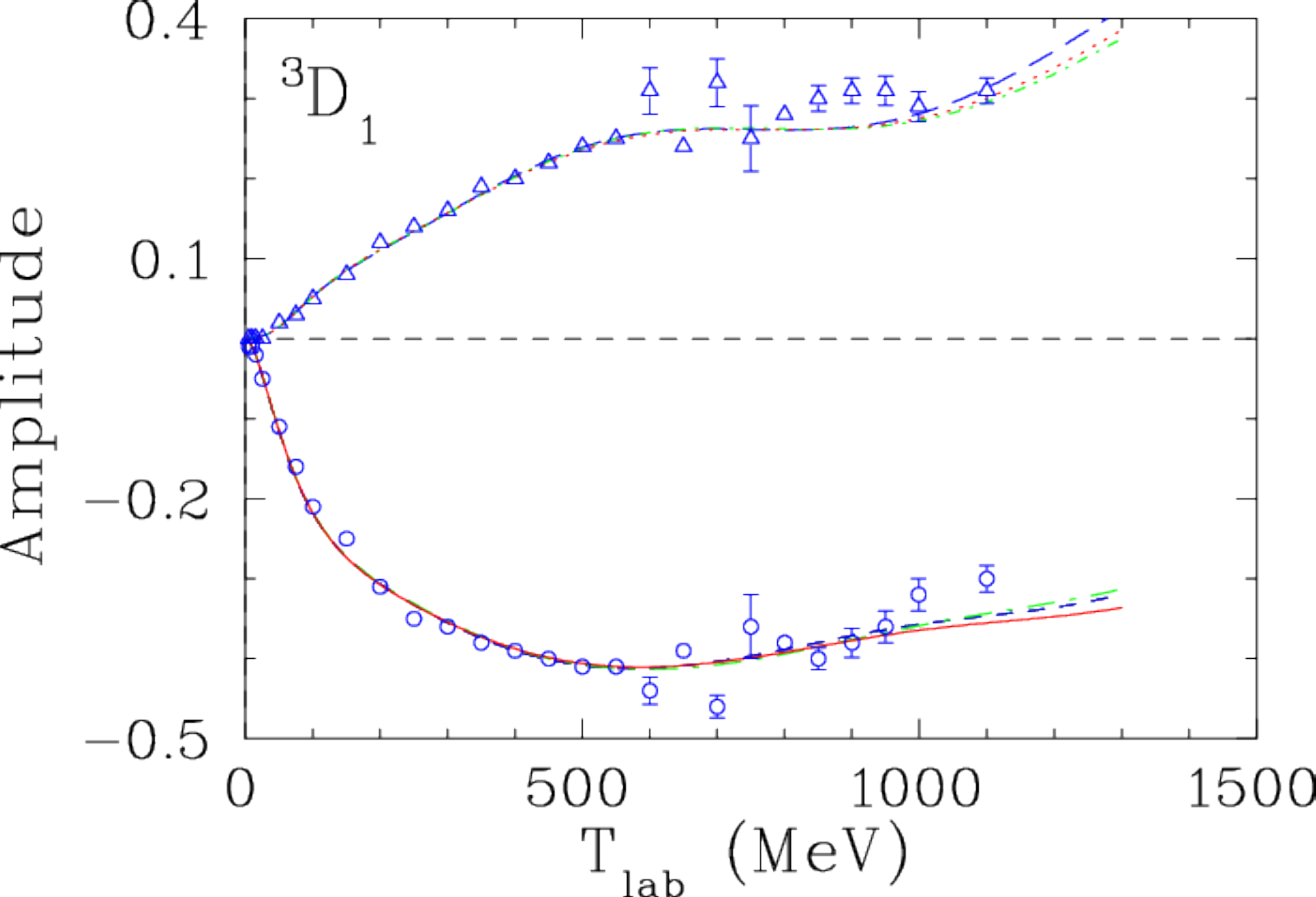}}
\centerline{
\includegraphics[height=0.45\textwidth, angle=90]{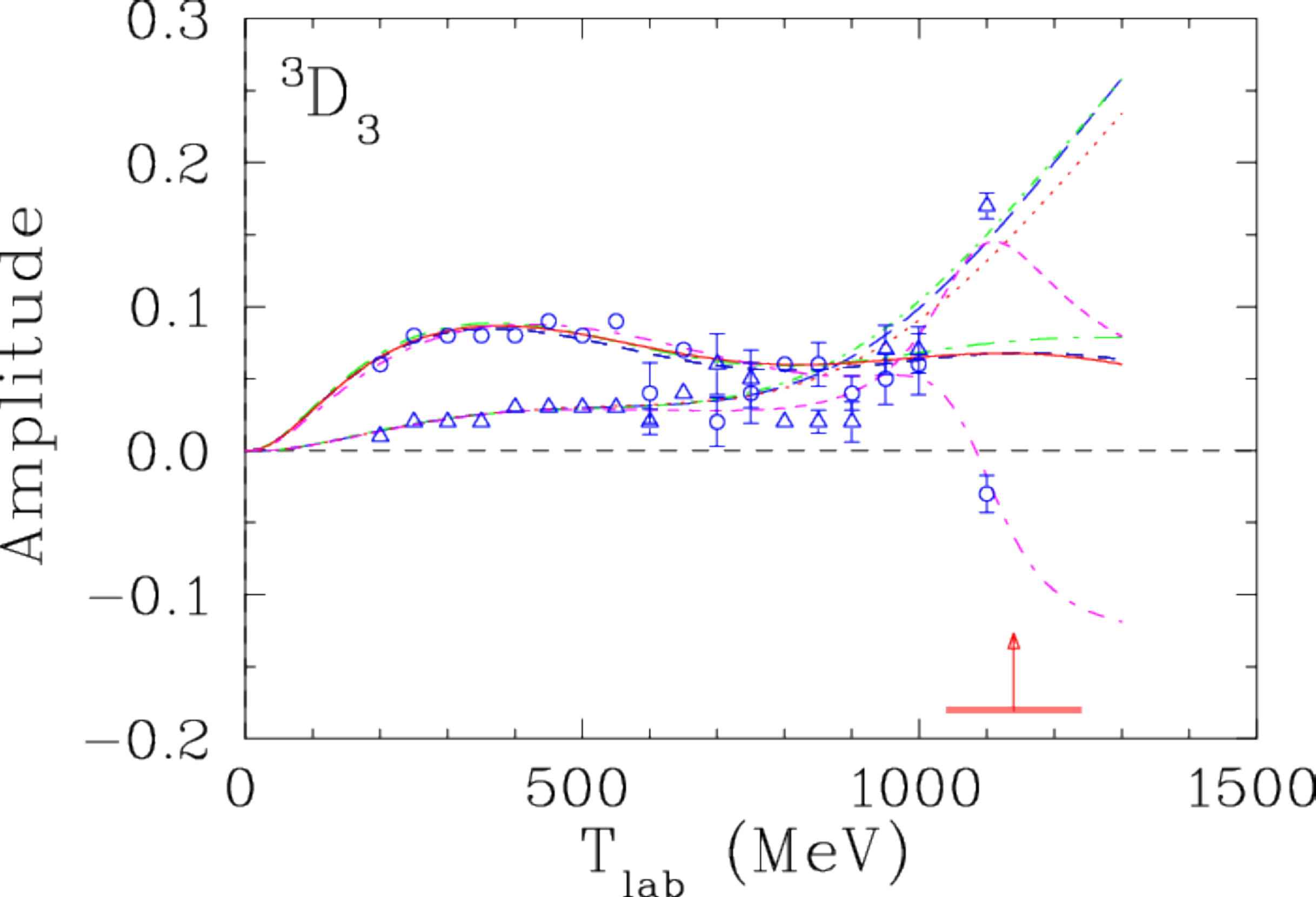}\hfill
\includegraphics[height=0.45\textwidth, angle=90]{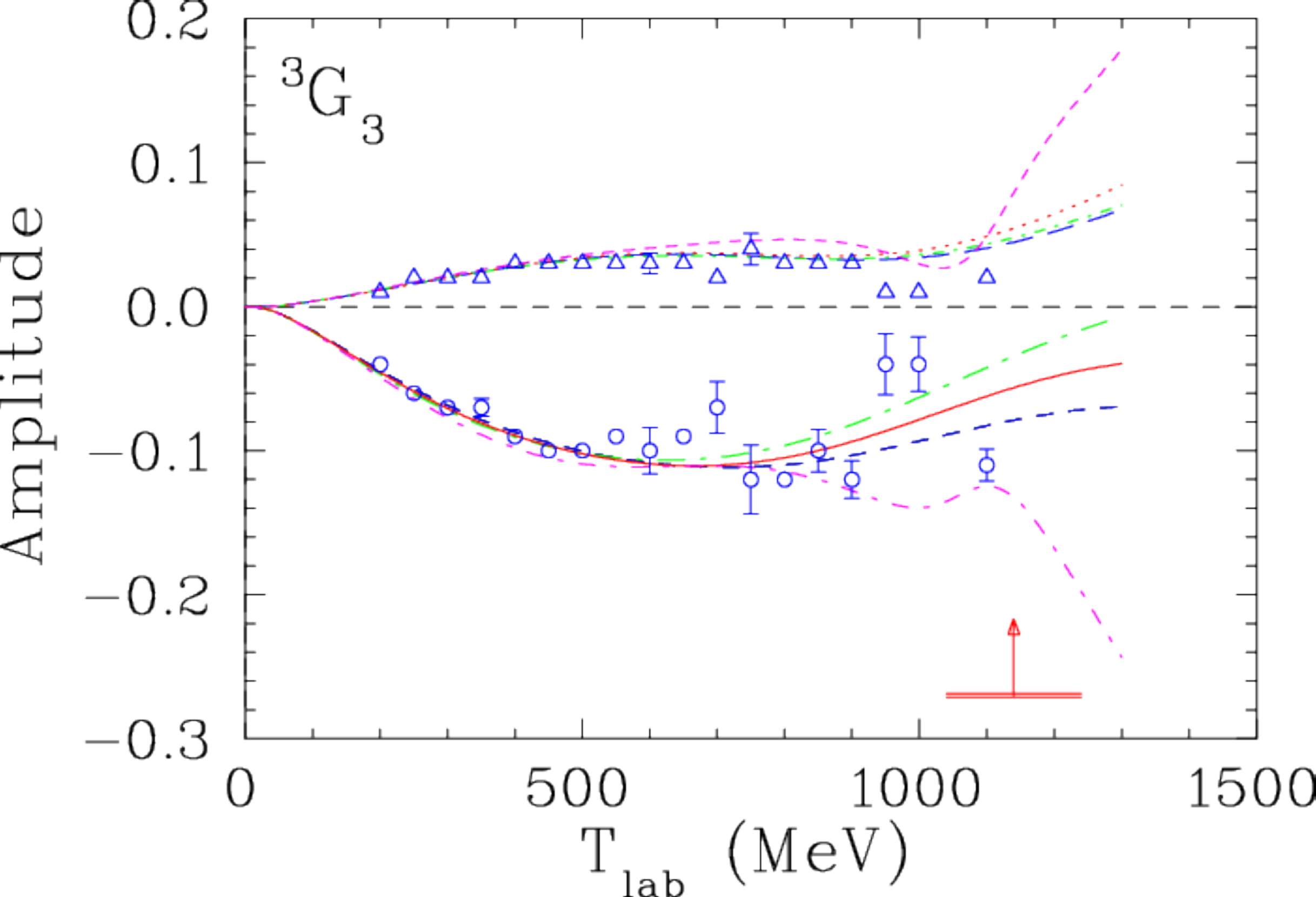}}
\centerline{
\includegraphics[height=0.45\textwidth, angle=90]{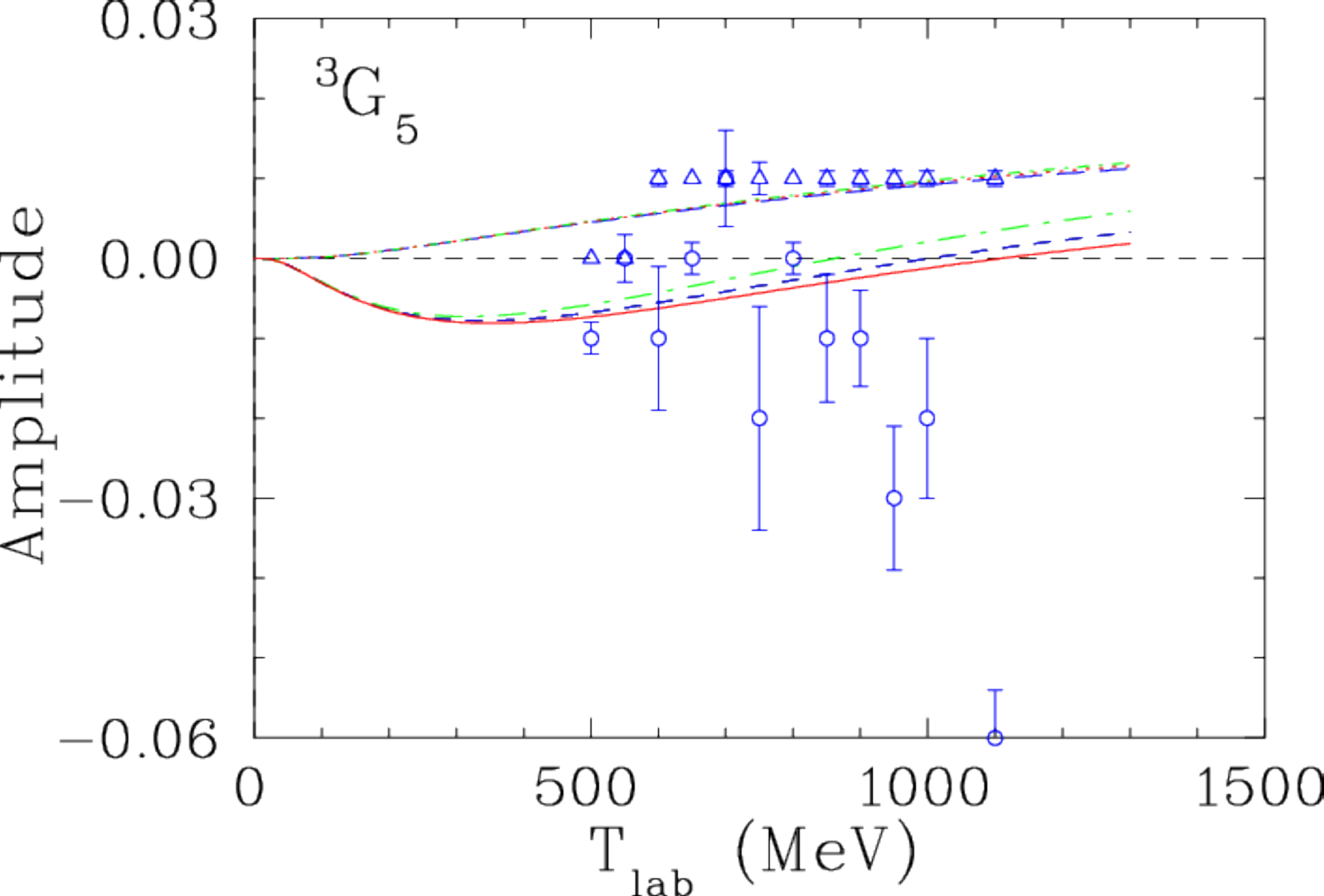}\hfill
\includegraphics[height=0.45\textwidth, angle=90]{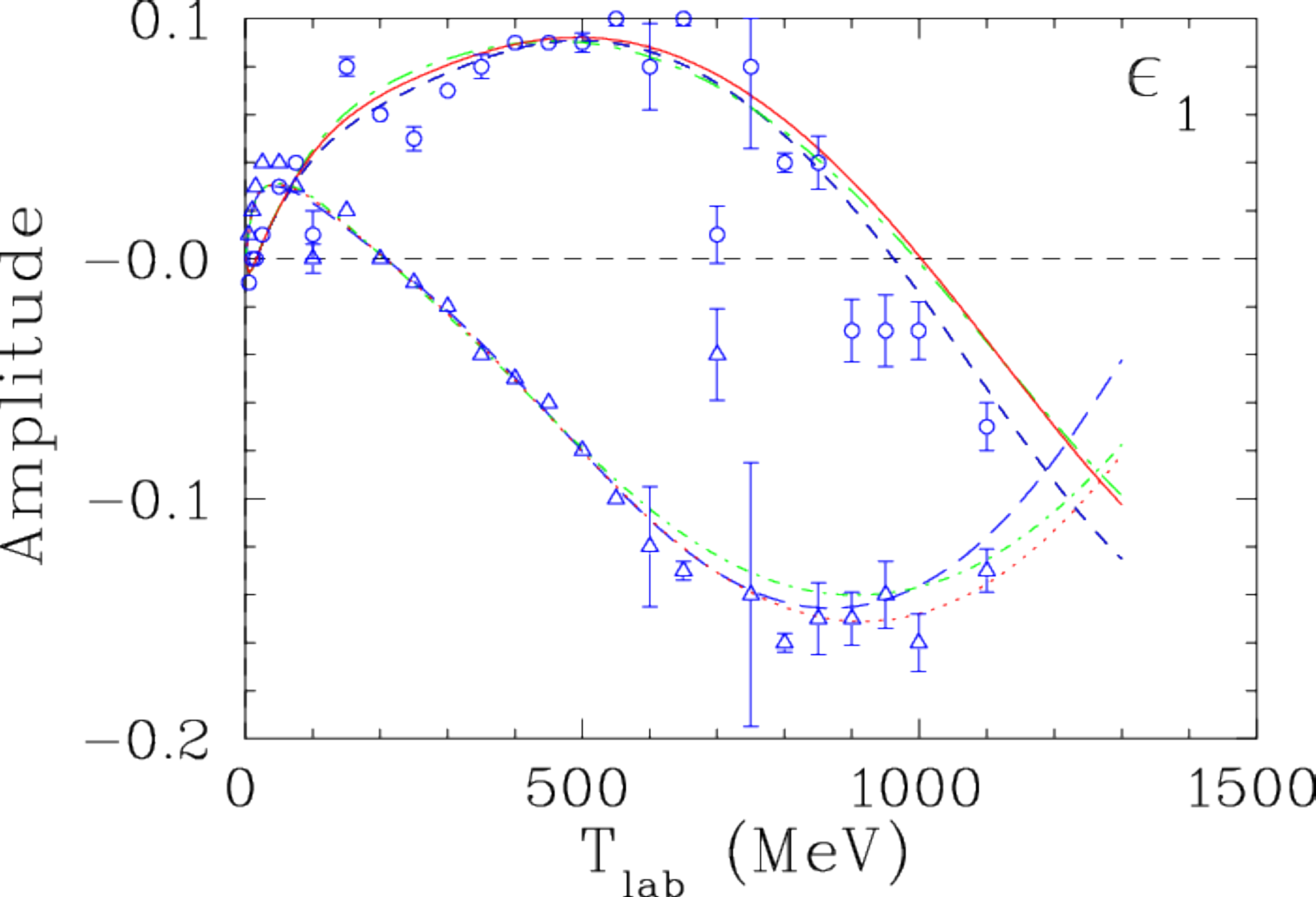}}
\centerline{
\includegraphics[height=0.45\textwidth, angle=90]{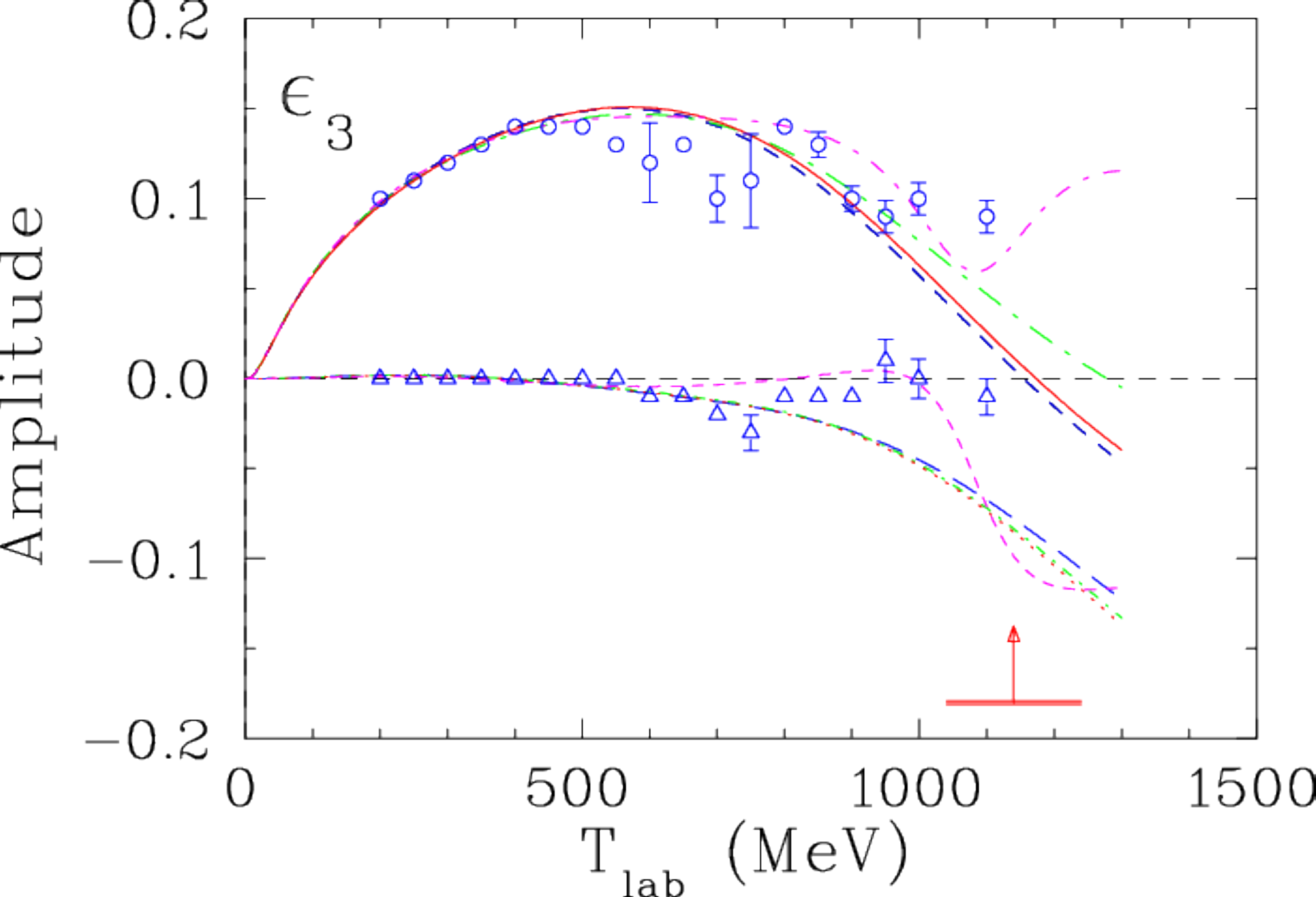}\hfill
\includegraphics[height=0.45\textwidth, angle=90]{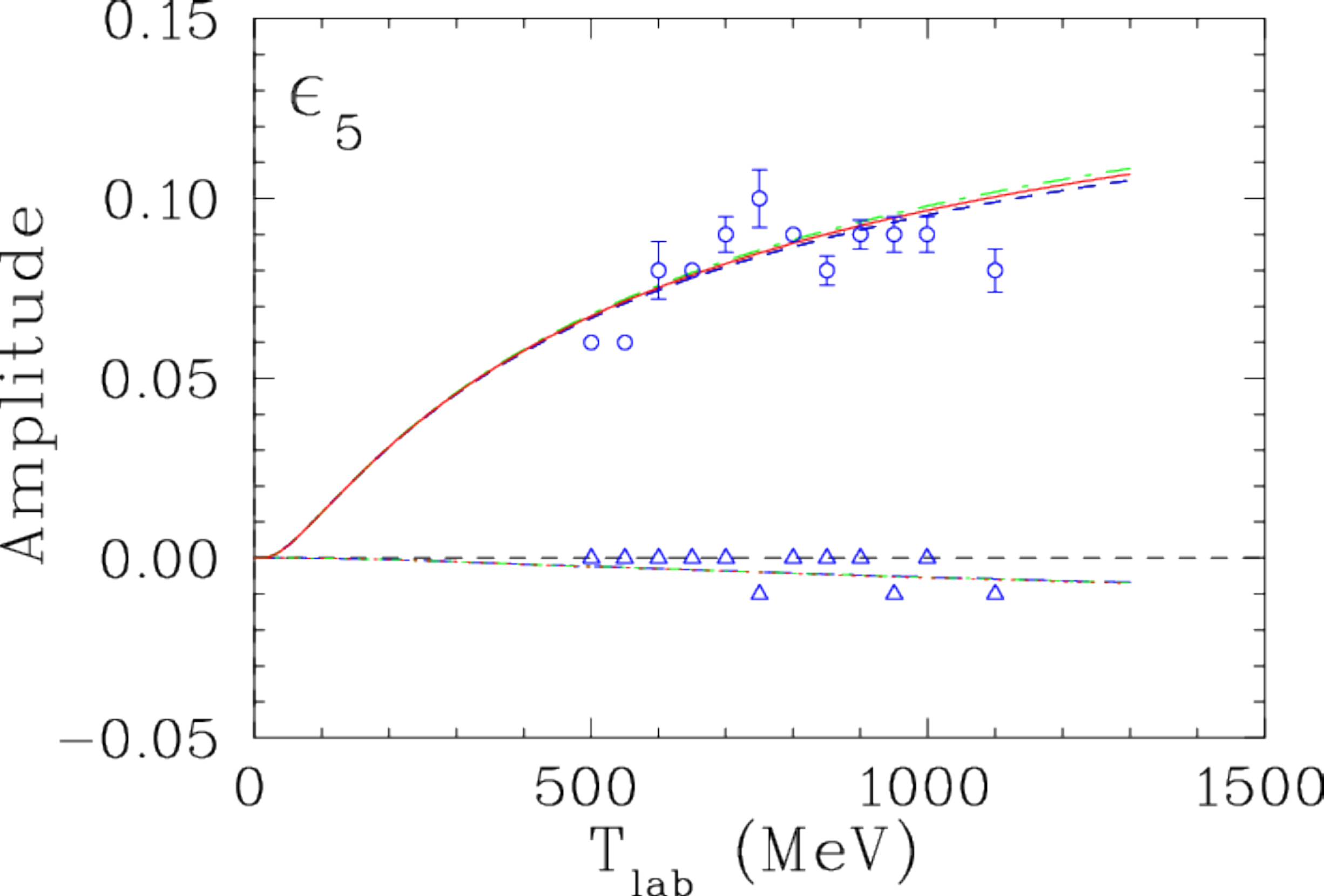}}
\vspace{3mm}
\caption{(Color online) Dominant isoscalar partial-wave
        amplitudes from threshold to $T_{lab}$ = 1.3~GeV.
        Notation as in Fig.~\protect\ref{fig:g11}.
        Revised SAID solution for $^3D_3$, $^3G_3$, and
        $\epsilon_3$, with a pole, is plotted as magenta 
	short dot-dashed (short dashed) lines for the real 
	(imaginary) parts. Red vertical arrows indicate the 
	WASA resonance mass $W_R$ value and the red horizontal 
	bar gives the full width
        $\Gamma$~\protect\cite{WASA}. \label{fig:g13}}
\end{figure*}

\begin{figure*}[th]
\centerline{
\includegraphics[height=0.45\textwidth, angle=90]{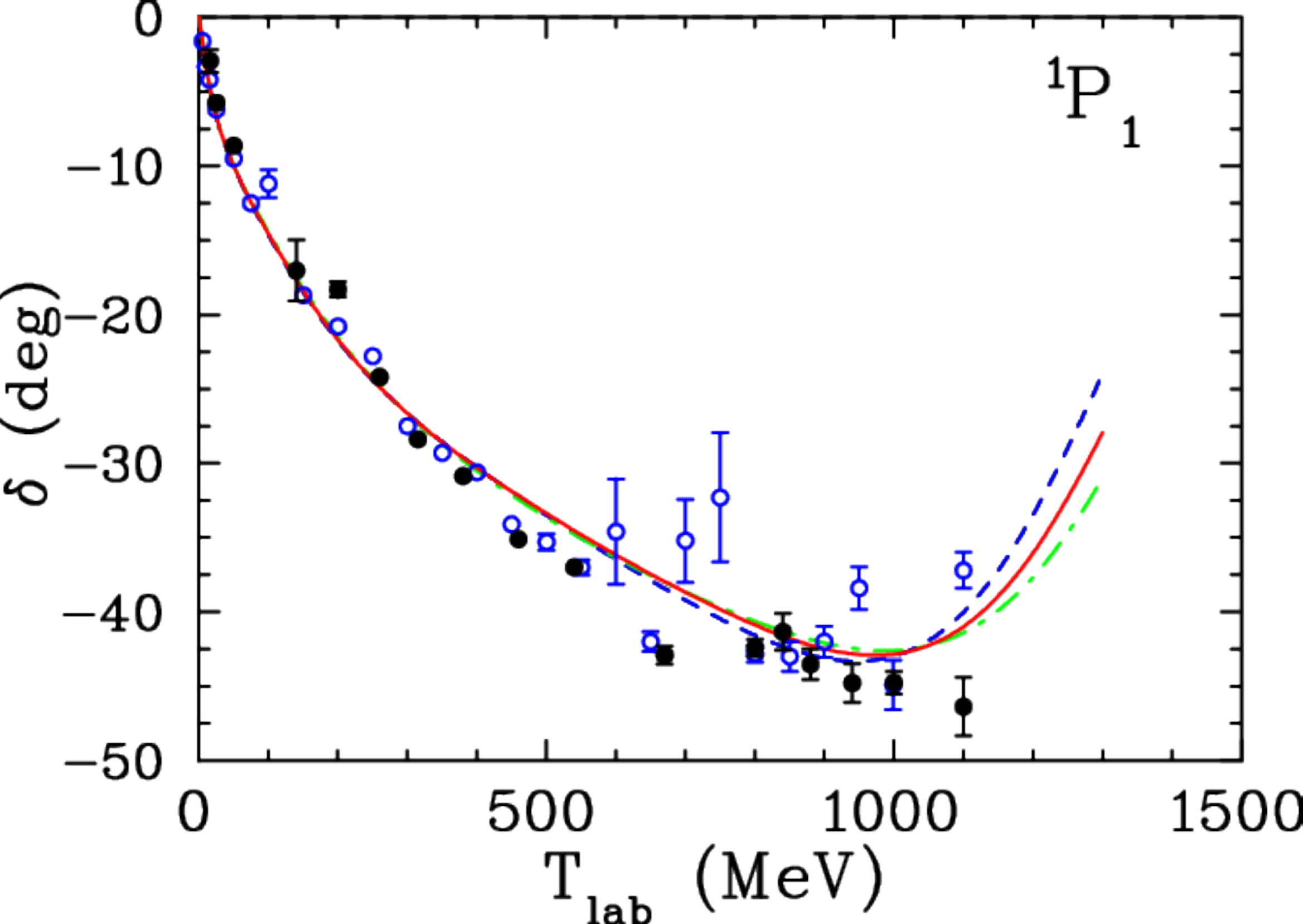}\hfill
\includegraphics[height=0.45\textwidth, angle=90]{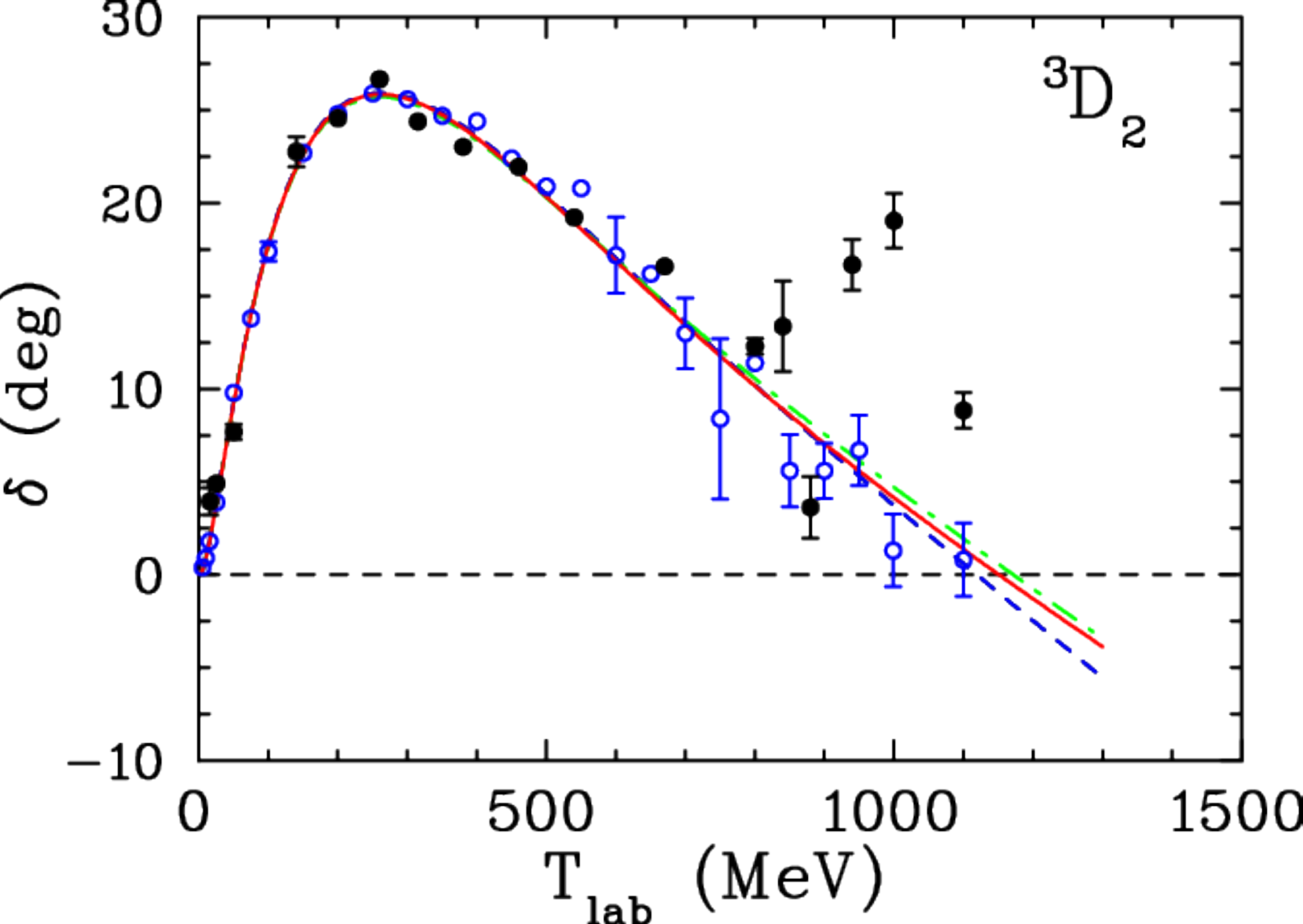}}
\centerline{
\includegraphics[height=0.45\textwidth, angle=90]{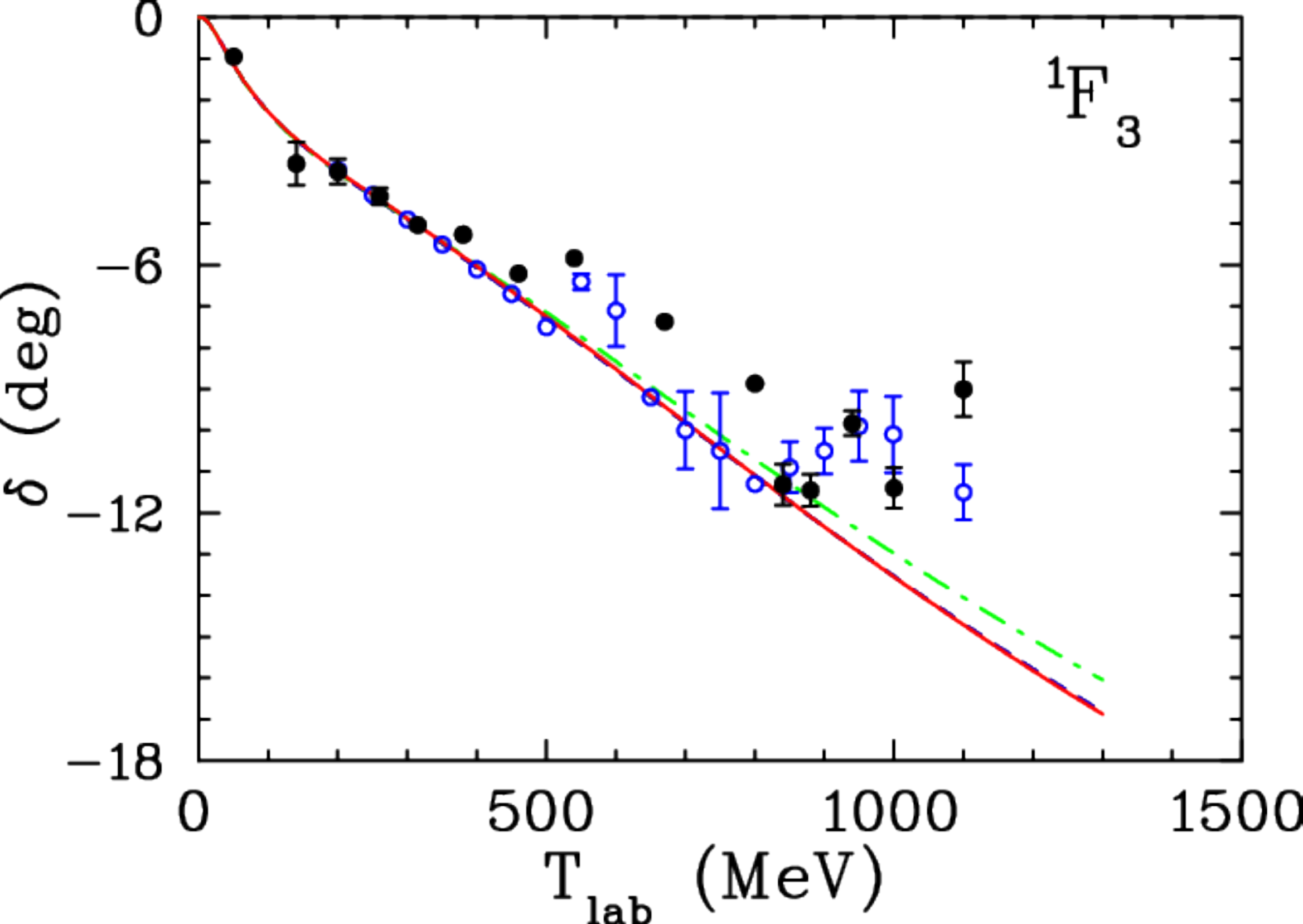}\hfill
\includegraphics[height=0.45\textwidth, angle=90]{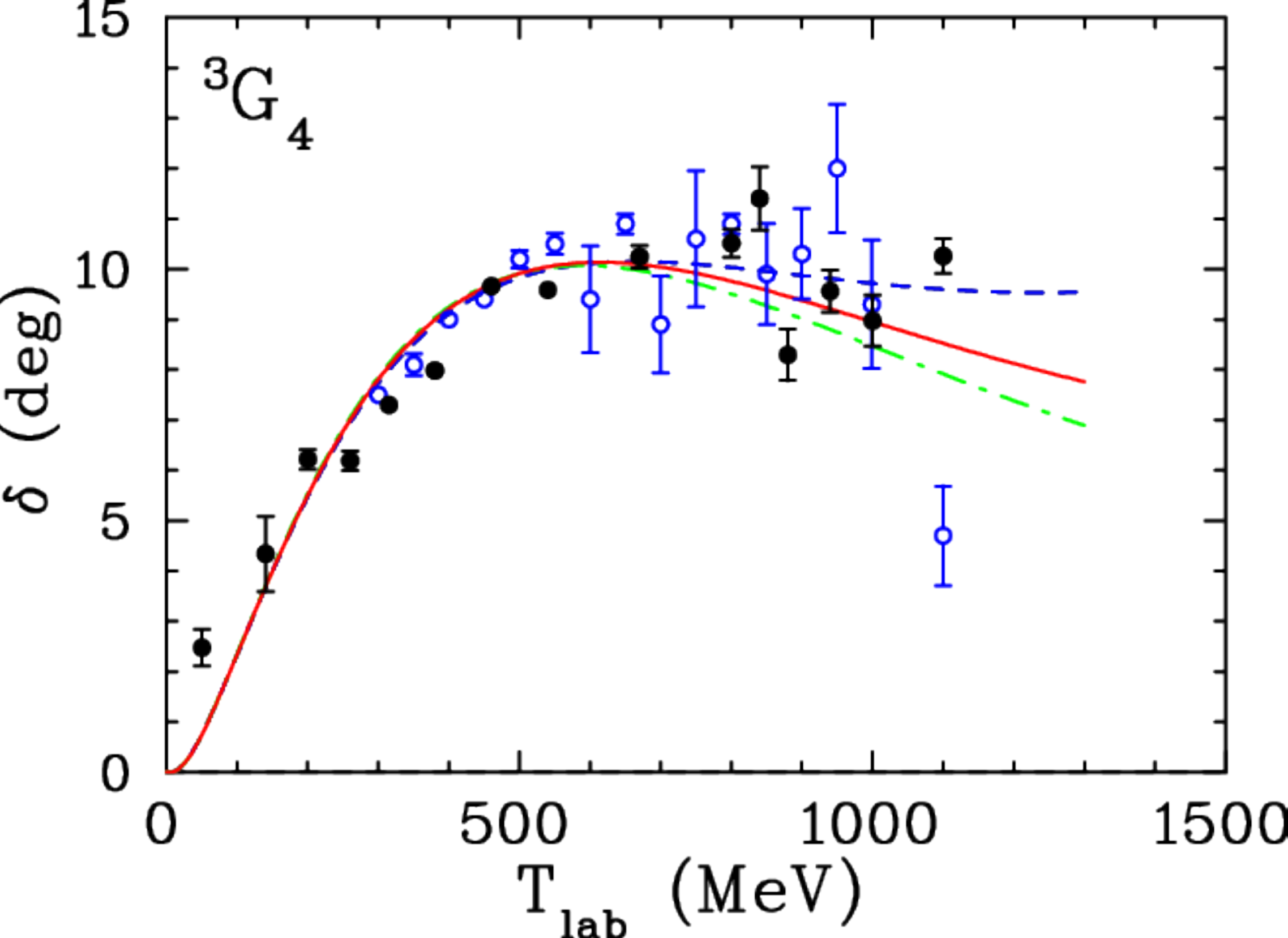}}
\vspace{3mm}
\caption{(Color online) Phase-shift parameters for dominant
        isoscalar partial-wave amplitudes from threshold to
        $T_{lab}$ = 1.3~GeV.  The Saclay SES~\protect\cite{Ball98} 
	are given by black filled circles; SAID SES are plotted
        as open blue circles.  Notation for SAID
        energy-dependent amplitudes 
	as in Fig.~\protect\ref{fig:g1}. \label{fig:g14}}
\end{figure*}

The dominant isoscalar partial-wave amplitudes are plotted in
Figs.~\ref{fig:g13} and \ref{fig:g14}. As for the isovector waves,
we first plot the global fits and SES in terms of the dimensionless
T-matrices. Here, for the coupled $^3D_3$--$^3G_3$ partial waves, we
also plot the fit containing an associated pole for comparison. This
feature has been extensively discussed in Refs.~\cite{WASA,dibaryon_GW}.
Phase shifts are plotted for selected waves in Fig.~\ref{fig:g14}. As
was the case for the isovector amplitudes, all determinations,
including the Saclay PWA~\cite{Ball98} agree well up to 500 MeV.

%%%%%%%%%%%%%%%%%%%%%%%%%%%%%%%%%%%%%%%%%%%%%%%%%%%
%%%%%%%%%%%%%%%%%%%%%%%%%%%%%%%%%%%%%%%%%%%%%%%%%%%%
%%%%%%%%%%%%%%%%%%%%%%%%%%%%%%%%%%%%%%%%%%%%%%%%%%%%
%%%%%%%%%%%%%%%%%%%%%%%%%%%%%%%%%%%%%%%%%%%%%%%%%%%%%
%%%%%%%%%%%%%%%%%%%%%%%%%%%%%%%%%%%%%%%%%%%%%%%%%%%%%%
%%%%%%%%%%%%%%%%%%%%%%%%%%%%%%%%%%%%%%%%%%%%%%%%%%%%
%%%%%%%%%%%%%%%%%%%%%%%%%%%%%%%%%%%%%%%%%%%%%%%%%%%%%
%%%%%%%%%%%%%%%%%%%%%%%%%%%%%%%%%%%%%%%%%%%%%%%%%%%%%%

\section{Summary and Conclusions}
\label{sec:summary}

We have fitted recent precise measurements of the
$pp$ differential cross sections and the observable $A_y$, both at forward
angles, together with $np$ $A_y$ data near the limit of our isoscalar
PWA. The energy dependence implied by the new $np$ data, together with older
measurements, suggest a rapid variation that has been accommodated through
the generation of a pole, providing supporting evidence for the WASA dibaryon
candidate. Little additional $np$ data has been added to the SAID database 
since the 2007 PWA and the sharp structure in $A_y$ is localized to a 
narrow energy range. As a result, the amplitudes - apart from the coupled
$^3D_3$--$^3G_3$ partial waves - have changed little since the last published
analysis. At the level of a DAR, we are consistent with the Saclay analysis.
However, the upper limit for our $np$ PWA (1.3 GeV) is much below the $pp$
PWA upper limit (3.0 GeV).  

The $pp$ DAR comparison has been made at a higher set of energies and here
multiple solutions appear to form branches that are chosen differently by
the present SM16 and WF16 fits. The SM16 and WF16 fits also differ at the
PWA level, particularly in the $^1S_0$, $^1D_2$, and $^3F_2$
amplitudes. The largest difference, in the $^1S_0$ wave, is very 
sensitive to the fit quality associated with
the polarized and unpolarized $pp$ total cross sections, 
together with the fit to near-forward $pp$ differential
cross sections. The present SM16 and WF16 solutions differ significantly from
SP07 in their fit to these observables. We have also observed that the
fit WF16 has a $^1S_0$ partial wave remarkably similar to that found in
the older SM97 solution, which featured a different treatment of the
total cross sections. 

The weighted fit, WF16, was constructed to improve the description of new
cross section and $A_y$ data. However, by increasing the weight of new data,
older data are implicitly given less weight, resulting in a degraded
fit. This produces a higher overall $\chi^2$/datum for the full
database. For values of ${\rm T_{lab}}$ from threshold to 350 MeV, the
potential model region, all fits are virtually identical, giving a 
$\chi^2$/datum of about 1.25 for $pp$ over the full database. 
For $pp$ scattering up to 1 GeV, the $\chi^2$/datum increases to about
1.5, compared to a value slightly below 2.0 for the SM16 fit covering the
full $pp$ scattering range from threshold to 3 GeV.

We note that the
Nijmegen partial-wave and potential model analyses~\cite{Nijmegen} 
of $pp$ scattering data achieved a $\chi^2$/datum closer to
unity over a more restricted dataset and energy range. 
The Nijmegen low-energy partial waves
differ very little from those presented here.
Comparisons of the various analyses, 
including the Nijmegen fit, may be obtained using the SAID website. 

%%%%%%%%%%%%%%%%%%%%%%%%%%%%%%%%%%%%%%%%%%%%%%%%%%%%
\begin{acknowledgments}
This work was supported in part by the U.S.\ Department of Energy,
Office of Science, Office of Nuclear Physics, under award numbers
DE-SC0014133 and DE-SC0016582.
\end{acknowledgments}
\eject
%%%%%%%%%%%%%%%%%%%%%%%%%%%%%%%%%%%%%%%%%%%%%%%%%%%%

\end{document}